\documentclass[twocolumn]{aastex701}
\usepackage{amsmath}     
\usepackage{wasysym}           
\usepackage{graphicx}
\usepackage{amssymb}
\usepackage{epstopdf}
\usepackage{mathrsfs}
\usepackage{anyfontsize}
\usepackage{natbib}
\usepackage{makecell}
\usepackage{enumitem}
\usepackage{color}
\usepackage{lipsum}
\usepackage{diagbox}

\begin{document}

\title{On the origin of anomalous dissipation in simulations of tidal disruption events}
\shorttitle{Anomalous dissipation in TDEs}
\shortauthors{Nixon et al.}
\correspondingauthor{C.~J.~Nixon}
\author[0000-0002-2137-4146]{C.~J.~Nixon}
\affiliation{School of Physics and Astronomy, University of Leeds, Sir William Henry Bragg Building, Woodhouse Ln., Leeds LS2 9JT, UK}
\email[show]{c.j.nixon@leeds.ac.uk}

\author[0000-0003-3765-6401]{Eric R.~Coughlin}
\affiliation{Department of Physics, Syracuse University, Syracuse, NY 13210, USA}
\email[show]{ecoughli@syr.edu}

\author[0000-0001-5064-1269]{Zachary L.~Andalman}
\affiliation{Department of Astrophysical Sciences, Princeton University, 4 Ivy Lane, 08540,Princeton, NJ, USA}
\email[]{zack.andalman@princeton.edu}

\begin{abstract}
In a tidal disruption event (TDE), a star is destroyed by the tidal field of a supermassive black hole. The stellar debris is initially placed on highly elliptical orbits, and a longstanding question in TDE theory is: How does the stellar debris circularize into a disc and accrete? The originally proposed answer to this question is self-intersection shocks, where relativistic apsidal precession results in a strong collision between the incoming and outgoing material. However, global simulations of TDEs tend to find enhanced hydrodynamical dissipation prior to any intersections of the debris orbits, with the material ``fanning out'' into a wide-angle and partially-unbound outflow upon passing through pericenter. We show that this dissipation is numerical in origin and arises from a combination of 1) the change in the kinematics of the debris as it passes through pericenter, with its velocity profile along the stream transitioning from strongly diverging pre-pericenter to strongly converging post-pericenter, and 2) the dependence of numerical algorithms (viscosity switches for particle-based methods and Riemann solvers for Godunov-based schemes) on the diverging vs.~converging nature of the fluid. We support this conclusion with analytical and numerical modeling. We discuss possible resolutions to these issues as well as the implications of our findings in the context of observations.
\end{abstract}



\section{Introduction}
Tidal disruption events (TDEs) occur when a star is torn apart by the tidal field of a massive black hole \citep{Hills:1975, Lacy:1982, Rees:1988, Gezari:2021}. The stellar debris that is bound to the black hole is placed onto highly eccentric orbits that -- through several poorly understood processes -- forms an accretion flow. Some of the accretion power likely generates outflows and some of it escapes as radiation (directly or reprocessed by surrounding gas) that we observe, as borne out by numerous observations (e.g., \citealt{Bloom:2011, Cenko:2012, Gezari:2012, Arcavi:2014, Alexander:2016, Kara:2016, Pasham:2018, Holoien:2019, Nicholl:2019, Nicholl:2020, Cendes:2021, Horesh:2021, Hung:2021, Sazonov:2021, vanVelzen:2021, Andreoini:2022, Goodwin:2023, Hammerstein:2023, Pasham:2023, Yao:2023, Christy:2024, Wevers:2024, Ajay:2025, Ho:2025, Sfaradi:2025, Yao:2025, Christy:2026, Hammerstein:2026}).

Our basic understanding of TDE dynamics was set forth by \citet{Hills:1975, Lacy:1982, Rees:1988} and is as follows: If the debris followed Keplerian ellipses without significant internal (i.e., fluid-dynamical) dissipation, an elliptical disc would form with a small range of pericentre distances and a large range of apocentre distances. However, as the pericentres are typically tens of Schwarzschild radii, the orbits undergo relativistic precession, leading the gas orbits to intersect and shock. These shocks circularize the flow by removing orbital energy but not angular momentum, and the accretion flow is then expected to proceed in the way envisaged in \cite{Shakura:1973} for super-Eddington flows, such that the mass flow rate on to the black hole is maintained at the Eddington limit with the remainder expelled in a disc wind.\footnote{\label{footnote:1}It is also possible that self-intersections lead to the ejection of material, i.e., some of the shocked material is unbound at the expense of more tightly binding other material \citep[][but see \citealt{Huang:2024}]{Jiang:2016, Bonnerot:2020}.}

\cite{Rees:1988} also discusses the possibility that the Eddington limit is obviated by the accretion of low-binding-energy, low-angular-momentum orbits that plunge $\sim$ directly into the black hole, thereby releasing effectively none of their orbital energy as they do so. How much stellar debris is directly accreted in this fashion, accreted through a disc-like flow, or expelled from the disc in a wind (or from circularization shocks; see Footnote~\ref{footnote:1}) likely depends sensitively on the parameters of the TDE, including the pericenter of the stellar orbit, the physical properties of the star, and the black hole mass and spin.

Since these original works, there have been many attempts to characterize these dependencies and understand the overall efficiency of circularization, many of which have been via three-dimensional hydrodynamical simulations. There are some aspects of the ``numerical TDE problem'' -- TDE dynamics analyzed with three-dimensional hydrodynamical simulations -- for which there is overall agreement in the literature, such that different codes with the same numerical methodology, and different codes with different numerical methodologies, yield the same answer. For example, the maximum radius at which a star is completely destroyed \citep[e.g.][]{Guillochon:2013,Gafton:2015,Mainetti:2017,Miles:2020,Coughlin:2022b} and the energy distribution of the debris when the star does not plunge too deeply within the tidal radius \citep[e.g.][]{Lodato:2009,Guillochon:2013,Norman:2021} are widely agreed upon.

But, there are aspects of the TDE problem that are known to present difficulties for numerical simulations, even at modern day resolution. For example, \cite{Norman:2021} and \citet{Coughlin:2022} showed that SPH simulations of deeply plunging TDEs can include substantial numerical error---even when 128 million particles are included---in the breadth of the debris energy distribution and the amount of thermal heating incurred by the gas.

Another example, which is the focus of the present investigation, is the hydrodynamics of the stream as it returns to pericenter: the debris stream expands predominantly radially and cools, connoting the idea that the circularization and disc formation is initiated by the collision of the geometrically thin, kinetic-energy-dominated outgoing stream with the (also thin and cold) incoming stream at effectively a single point. However, while the dynamics of the incoming stream have been found (numerically) to reflect this set of notions and analytical predictions thereof \citep{Kochanek:1994, Coughlin:2016}, both finite-mass and finite-volume methods have found that the returning stream is substantially thermodynamically and kinematically modified (e.g., \citealt{Lee:1996, Ayal:2000, Bogdanovic:2004, Guillochon:2014, Shiokawa:2015, Ryu:2023, Steinberg:2024, Price:2024}) upon passing through pericenter: the debris spreads laterally into a wide-angle ``fan,'' it is significantly hotter, and a fraction of it (\citealt{Ayal:2000} suggested as much as 75\% of the material initially bound to the black hole, or $\sim 37\%$ of the material overall) is ejected on unbound orbits.

There have been suggestions that all of these effects are numerical in origin \citep{Bonnerot:2022, Huang:2024}, and that their magnitude depends on resolution has been directly demonstrated with Smoothed Particle Hydrodynamics (SPH). For example, \citet{Kubli:2025} presented a sequence of simulations ranging from $10^6$ to $10^{10}$ particles, studying the return of the material from a $5/3$-polytrope (with a solar mass and radius) being destroyed a $10^6 M_{\odot}$ black hole (the same setup used in, e.g., \citealt{Bicknell:1983, Evans:1989, Ayal:2000}). They found that the anomalous fanning, or ``spraying,'' of the material as it returns to pericenter is effectively absent at $10^{10}$ particles. The same authors also showed that the total amount of dissipation, i.e., the fraction of kinetic energy dissipated through numerical viscosity, was converging in that it declined monotonically with increasing resolution, but had not yet converged.

Despite these strong intimations of the non-physical nature of the anomalous dissipation recovered by numerical methods, the precise reasons as to where and why these methods fall short---both that SPH and grid-based methods have struggled to achieve converged stream properties near pericentre at extreme resolution---have not been identified. It is our intent here to try to elucidate, or at least shed some light on, the origin of the numerical dissipation found in simulations of TDEs. In Section \ref{sec:analytic} we develop an analytical model, which combines past approaches and uses both Eulerian and Lagrangian techniques, to understand the hydrodynamic properties of the stream as it returns to the black hole. In Section~\ref{sec:Origin} we use this model alongside hydrodynamical simulations to assess the origin of the fanning and heating of the debris as it returns, which we argue stems primarily from the diverging-to-converging and highly supersonic nature of the velocity field along the stream near pericenter, coupled with the dependence of fluid-dynamical algorithms (specifically those for handling multi-valued velocity fields) on these aspects of the flow. We also assess the accuracy of these hypotheses with SPH simulations, and show that when the $\beta$-viscosity term is excluded from the SPH algorithm, there is no spraying of the material. In Section \ref{sec:discussion} we discuss potential solutions to these issues and approaches that have been tried but lead to unphysical results, such as the particle splitting method (\citealt{Ayal:2000} and \citealt{Hu:2026}, see also Appendix \ref{sec:appA}), and we also demonstrate that both outcomes envisaged by \citet{Rees:1988}---quasi-circular accretion vs.~near-parabolic infall---can be recovered from SPH simulations, depending on the particle number and the choice of viscosity parameters. We summarize in Section \ref{sec:summary}.

\section{Analytic model of the debris hydrodynamics}
\label{sec:analytic}
The stream of debris resulting from a TDE evolves non-trivially, owing to the continued post-disruption influence of pressure and self-gravity. This aspect of the problem was highlighted by \citet{Kochanek:1994}, who used a Lagrangian approach alongside an affine prescription for the transverse structure of the stream to study its dynamics under the influence of gas pressure, self-gravity, hydrogen recombination, viscous effects, and relativistic gravity. This same general methodology has since been used by other authors \citep{Bonnerot:2022b, Andalman:2025}. In contrast, it was pointed out in \citet{Coughlin:2016} that, since all the debris originates very near the marginally bound (i.e., zero-Keplerian-energy) radius, one can accurately treat the Eulerian velocity along the stream self-similarly, such that the dynamics depend only on time relative to the local dynamical time. They then showed that the Eulerian density profile can be recovered by considering the behavior of the stream width, the evolution of which changes qualitatively in the bound vs.~unbound segments of the stream. 

To understand the stream evolution, here we combine these two approaches: we treat the stream as extending purely radially from the black hole\footnote{More generally, one can define the radial direction as a curvilinear axis that extends in the direction of the maximum-density curve, and curvature and centrifugal terms that incorporate the finite angular momentum of the debris remain small until the stream returns to near pericenter; see \citet{Coughlin:2023} for more details.} and assume that the motion along the stream is purely kinematic and dominated by the black hole's gravitational field. Then from \citet{Coughlin:2016} we know that the velocity field can be described self-similarly, i.e., we write
\begin{equation}
    v = V(t) f\left(\xi\right), \,\,\, \xi = \frac{r}{R(t)}, \,\,\, V(t) = \frac{dR}{dt} = \sqrt{\frac{2GM_{\bullet}}{R(t)}}, \label{ssdef}
\end{equation}
where $v$ is the Eulerian radial velocity, $M_{\bullet}$ is the black hole mass, and $R(t)$ is the marginally bound radius that satisfies (from the right-most equality in Equation \ref{ssdef})
\begin{equation}
    R(t) = R_{\rm i}\left(1+\frac{3}{2}\frac{\sqrt{2GM_{\bullet}}}{{R_{\rm i}^{3/2}}}t\right)^{2/3} \label{Roft}
\end{equation}
with $R_{\rm i}$ some scale radius that is comparable to the tidal radius. This states that the velocity depends only on the relative distance from the zero-energy-Keplerian-orbit, $R(t)$, which is a good approximation if the stream starts out relatively localized (which it obviously does; see \citealt{Coughlin:2016} for a comparison between this solution and hydrodynamical simulations). Then the equation for $f(\xi)$ that comes from the radial momentum equation is
\begin{equation}
    \left(f-\xi\right)\frac{df}{d\xi} = \frac{1}{2}\left(f-\frac{1}{\xi^2}\right). \label{feq}
\end{equation}
We require the continuity of the velocity and its derivative at the marginally bound radius; the former yields $f(1) = 1$. This boundary condition establishes a critical point in Equation \eqref{feq}, and using L'Hopital's rule and the fact that the velocity should increase with increasing distance from the black hole (i.e., unbound material is moving faster than bound) shows
\begin{equation}
    \frac{df}{d\xi}\bigg{|}_{\xi = 1} = 2.
\end{equation}
These boundary conditions are sufficient to numerically integrate Equation \eqref{feq}.

To establish the other properties of the stream, such as its width and density profile (with both distance along and perpendicular to the stream axis), it is more convenient to adopt a Lagrangian description. The equation of motion along the stream is then
\begin{equation}
\begin{split}
    &\frac{\partial r}{\partial t}\bigg{|}_{r_0} =  V(t)f\left(\xi\right) \\ 
    \Rightarrow \quad &\frac{\partial \xi}{\partial \tau}\bigg{|}_{\xi_0} = f-\xi, \,\,\, \tau = \ln\left(\frac{R(t)}{R_{\rm i}}\right). \label{xieq}
\end{split}
\end{equation}
Here $\xi_0 = r_0/R_{\rm i}$ is the initial radial coordinate of a fluid element relative to $R_{\rm i}$, which satisfies $0.99 \lesssim \xi_0 \lesssim 1.01$ for typical TDEs. We let the stream adopt a cylindrical cross-section and assume the Lagrangian position $r$ is independent of the initial cylindrical distance $s_0$, which effectively removes the possibility of circulation in the stream. We could relax this assumption and model the latter effect perturbatively, but we expect its impact to be small: bulk motion along the stream (i.e., independent of initial cylindrical radius) should primarily establish its hydrodynamical evolution. With these assumptions, the density is given by
\begin{equation}
    \rho(s_0,r_0,t) = \rho_0(s_0,z_0)\left(\frac{s}{s_0}\right)^{-1}\left(\frac{\partial s}{\partial s_0}\right)^{-1}\left(\frac{\partial r}{\partial r_0}\right)^{-1},
\end{equation}
where $\rho_0$ is the initial stream density profile. We will also assume that the stream behaves adiabatically,\footnote{Radiative recombination heats the stream, moves fluid elements to different adiabats, and can increase the stream thickness relative to the adiabatic value by a factor of a few \citep{Kochanek:1994, Kasen:2010, Coughlin:2016, Coughlin:2023, Steinberg:2024, Andalman:2025}, but these effects do not alter our conclusions.} in which case the pressure is established by entropy conservation:
\begin{equation}
    \begin{split}
    p(s_0,r_0,t) &= p_0(s_0,r_0)\left(\frac{\rho}{\rho_0}\right)^{\gamma} \\
    &= p_0(s_0,r_0)\left(\frac{s}{s_0}\right)^{-\gamma}\left(\frac{\partial s}{\partial s_0}\right)^{-\gamma}\left(\frac{\partial r}{\partial r_0}\right)^{-\gamma},
    \end{split}
\end{equation}
where $p_0$ is the initial pressure profile and $\gamma$ is the adiabatic index of the gas.

Barring the assumption of the independence of $r$ on $s_0$, the preceding expressions for the density and pressure are general, and must be coupled to the Poisson and cylindrical-radial momentum equations. We additionally assume that the motion is homologous in the cylindrical-radial direction, such that
\begin{equation}
    s(s_0,r_0,t) = H(r_0,t) s_0,
\end{equation}
where $H(r_0,t)$ is a function that is determined from the momentum equation in the cylindrical-radial direction. We know that this parameterization is self-consistent near the marginally bound radius (see Appendix A of \citealt{Coughlin:2023}) and captures the bulk evolution of the stream. Then the density and pressure are
\begin{equation}
    \begin{split}
    \rho &= \rho_0(s_0,r_0)H^{-2}\left(\frac{\partial r}{\partial r_0}\right)^{-1}, \\
    p &= p_0(s_0,r_0)H^{-2\gamma}\left(\frac{\partial r}{\partial r_0}\right)^{-\gamma}.
    \end{split}
\end{equation}

The cylindrical-radial equation of motion contains gradients in terms of $s$. Because we are not allowing for circulation in the stream, it follows that
\begin{equation}
    \frac{\partial r_0}{\partial r} = \left(\frac{\partial r}{\partial r_0}\right)^{-1}, \quad \frac{\partial s_0}{\partial s} = \left(\frac{\partial s}{\partial s_0}\right)^{-1} = \frac{1}{H},
\end{equation}
and hence
\begin{equation}
    \frac{\partial}{\partial s} = \frac{\partial s_0}{\partial s}\frac{\partial}{\partial s_0} = \frac{1}{H}\frac{\partial }{\partial s_0}.
\end{equation}
The gravitational potential contains contributions from the black hole and the gas, where the former is
\begin{equation}
    \Phi_{\bullet} = -\frac{GM_{\bullet}}{\sqrt{s^2+r^2}} \quad \Rightarrow \quad \frac{\partial\Phi_{\bullet}}{\partial s} \simeq \frac{GM_{\bullet} s_0 H}{r^3},
\end{equation}
and the final equality ignores terms of the order $\sim s^3/r^3 \ll 1$. We will also let the initial density profile be uniform for simplicity,\footnote{The uniform-density assumption can be trivially relaxed under the present set of approximations, but the precise density profile depends on the early non-self-similar stream evolution, the finite angular momentum, and self-gravity along the stream, and the uniform-density solution is sufficient for our purposes here; see the discussion at the end of this section and the simulations in Section \ref{sec:simulations} that address the importance of the variation in the density along the stream.} such that $\rho_0(s_0,r_0) = \rho_{\rm c}$, in which case the self-gravitational potential satisfies
\begin{equation}
    \frac{\partial \Phi}{\partial s} = 2\pi G \rho_{\rm c} H^{-1}\left(\frac{\partial r}{\partial r_0}\right)^{-1}s_0.
\end{equation}
Finally, we assume that the initial pressure profile across the stream is
\begin{equation}
    p_0(s_0,r_0) = p_{\rm c}\left(1-\frac{s_0^2}{H_{\rm i}^2}\right),
\end{equation}
where $H_{\rm i}$ is the initial stream width and the quadratic term is necessary to balance the self-gravitational force (i.e., this is the leading-order in $s_0$, self-consistent form for the pressure profile across the stream; cf.~\citealt{Coughlin:2022}). 

With all of this, the equation for $H$ that follows self-consistently (i.e., such that all dependence on $s_0$ cancels) is
\begin{multline}
    \frac{\partial^2H}{\partial t^2}-\frac{2 p_{\rm c}}{\rho_{\rm c}H_{\rm i}^2}H^{1-2\gamma}\left(\frac{\partial r}{\partial r_0}\right)^{1-\gamma}  \\
    = -\frac{GM_{\bullet} H}{r^3}-2\pi G\rho_{\rm c}H^{-1}\left(\frac{\partial r}{\partial r_0}\right)^{-1}.
\end{multline}
The first term on the left-hand side is the Lagrangian acceleration of $H$ (i.e., $\partial/\partial t$ is the Lagrangian time derivative), the second term on the left-hand side arises from the pressure gradient, the first term on the right-hand side is the tidal potential of the black hole, and the second term on the right originates from the self-gravitational field of the stream. $H(r_0,t)$ is dimensionless and satisfies $H(r_0, t=0) = 1$, i.e., the preceding equation is dimensionally correct. Given that the motion along the stream was written in terms of $\xi$ and $\tau$, it is reasonable to use these same coordinates in our equation for $H$, which becomes
\begin{multline}
    2\ddot{H}-3\dot{H}-\frac{2p_{\rm c}R_{\rm i}^3}{\rho_{\rm c}H_{\rm i}^2GM_{\bullet}}H^{1-2\gamma}e^{\left(4-\gamma\right)\tau}\left(\frac{\partial\xi}{\partial \xi_0}\right)^{1-\gamma} \\
    = -\frac{H}{\xi^3}-\frac{2\pi \rho_{\rm c}R_{\rm i}^3}{M_{\bullet}}e^{2\tau}H^{-1}\left(\frac{\partial \xi}{\partial \xi_0}\right)^{-1}. \label{Heq}
\end{multline}
Here dots denote differentiation with respect to $\tau$.

To integrate Equation \eqref{Heq} we also require an initial condition for the transverse velocity, i.e., $\dot{H}(0)$. To this end, we note that there is an exact, ``equilibrium'' solution to Equation \eqref{Heq} near the marginally bound radius, where we can Taylor expand $\xi$ as $\xi = 1+\xi_0 e^{\tau}$ (note that this follows self-consistently from Equation \ref{xieq}, and $\xi_0 \lesssim R_{\star}/r_{\rm t} \lesssim 0.01$, i.e., this is highly accurate at early times), which is
\begin{equation}
    \gamma = 5/3, \quad \frac{p_{\rm c}}{\pi G \rho_{\rm c}^2 H_{\rm i}^2} = 1, \quad H(\xi_0,\tau) = e^{\tau/2}. \label{Hex}
\end{equation}
Note that $\gamma = 5/3$ should be accurate before and, to a lesser extent, after hydrogen recombination \citep{Andalman:2025}, and the second relation is effectively the one satisfied by hydrostatic balance in the original star, i.e., these conditions should automatically be upheld to high accuracy (this equilibrium solution was described in \citealt{Coughlin:2023}, and is the one about which their perturbation analysis was performed). Equation \eqref{Hex}, and the fact that it should be approximately satisfied in general, then motivates the following simplifications and initial conditions: let $\gamma = 5/3$, let $p_{\rm c}$, $\rho_{\rm c}$, and $H_{\rm i}$ be related via Equation \eqref{Hex}, and let $H(0) = 1$ and $\dot{H}(0) = 1/2$, implying that the stream begins in quasi-hydrostatic balance effectively everywhere (i.e., when the stream is still highly localized around the marginally bound orbit). With this set of conditions the equation for $H$ is
\begin{multline}
2\ddot{H}-3\dot{H}+\frac{H}{\xi^3}  \\ 
= \mu\left(-e^{2\tau}H^{-1}\left(\frac{\partial\xi}{\partial \xi_0}\right)^{-1} + H^{-7/3}e^{7\tau/3}\left(\frac{\partial\xi}{\partial \xi_0}\right)^{-2/3}\right), \label{Hfin}
\end{multline}
where
\begin{equation}
    \mu = \frac{2\pi \rho_{\rm c}R_{\rm i}^3}{M_{\bullet}}
\end{equation}
is the ratio of the stream density to the ``black hole density'' $M_{\bullet}/R_{\rm i}^3$ at the initial marginally bound radius. Because the fluid elements in the bound segment of the stream eventually start to return to the black hole, the stream width $H(\xi_0,\tau)$ will evolve in a way that is not simply given by $e^{\tau/2}$, but that can nonetheless be determined from the numerical solution to Equation \eqref{Hfin}, given $\xi(\xi_0,\tau)$ from Equation \eqref{xieq}. 

From the solution for $H(\xi_0,\tau)$ we can assess its Mach number and causal connectedness: the sound-crossing time over the stream width is given approximately by
\begin{equation}
    t_{\rm sc} = \frac{2H_{\rm i} H}{c_{\rm s}} = \frac{2}{\sqrt{\gamma \pi G \rho_{\rm c}}}H^{5/3}e^{\tau/3}\left(\frac{\partial \xi}{\partial \xi_0}\right)^{1/3},
\end{equation}
where $\tau$ corresponds to the position of the marginally bound radius when we impose some perturbation to the stream (note that the dimensionless stream width $H$ and $\partial\xi/\partial\xi_0$ are also evaluated at $\tau$) and recall that $H_{\rm i}$ is the initial stream width. The causal connectedness of the stream as concerns its ability to smooth out perturbations is established by the ratio of $t_{\rm sc}$ to the time taken for the fluid element to return to the origin $t_{\rm ret}$, which is 
\begin{equation}
    \frac{t_{\rm sc}}{t_{\rm ret}} \simeq 
\frac{H^{5/3}e^{\tau/3}\left(\frac{\partial \xi}{\partial \xi_0}\right)^{1/3}}{e^{3\tau/2}-e^{3\tau_1/2}}, \label{tsc}
\end{equation}
where $\tau_1$ is the position of the marginally bound radius when the fluid element reaches the origin, measured from $\tau = 0$. If this ratio is greater than one, the stream is causally disconnected transversely. 

\begin{figure*}
    \includegraphics[width=0.495\textwidth]{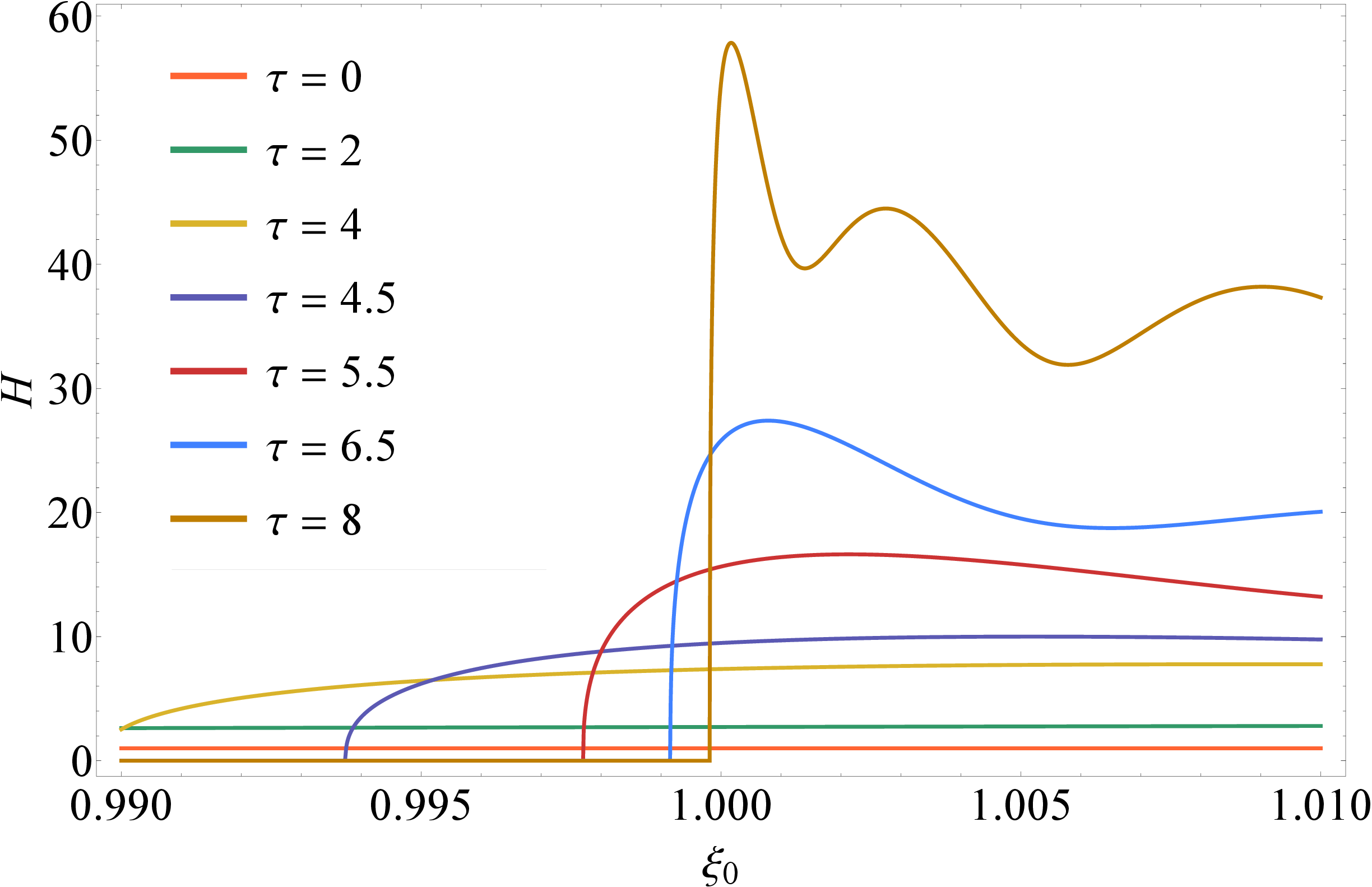}
    \includegraphics[width=0.495\textwidth]{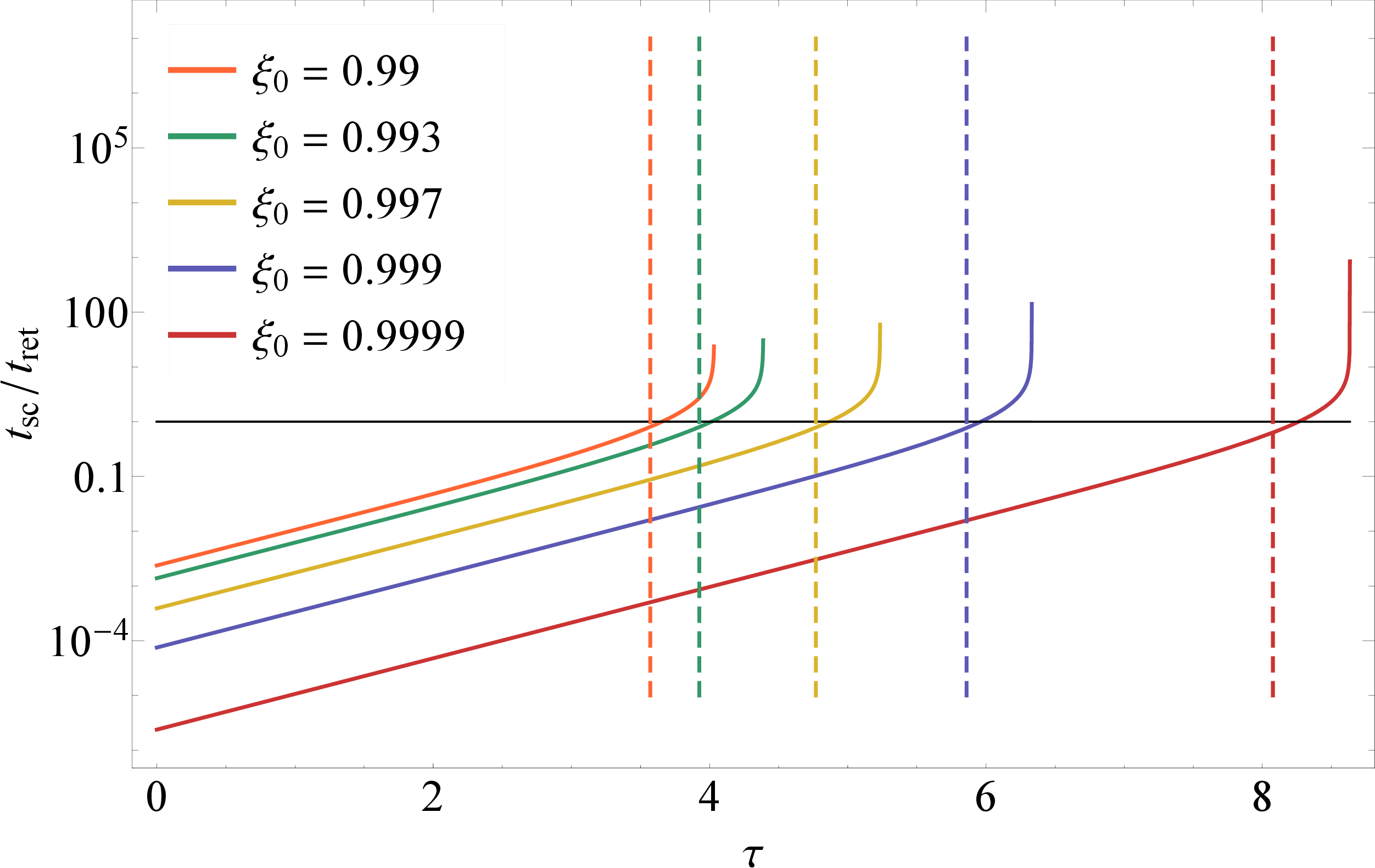}
    \caption{Left: The stream width, which in this set of approximations is the cross-sectional radius, as a function of initial (dimensionless) Lagrangian position $\xi_0$ evaluated at the dimensionless times $\tau$ (recall $\tau = \ln R(t)/R_{\rm i} \propto \ln t$) in the legend. The early-time behavior is characterized by a near-uniform increase at a rate that scales as $\propto e^{\tau/2}$, while at later times the bound fluid elements, which have $\xi_0 < 1$, return; the curves are stopped at the $\tau$ when $H = 1$, as at this point the stream has returned to pericenter and the finite angular momentum resists further compression.  The unbound fluid elements, which have $\xi_0 > 1$, show non-monotonic and oscillatory behavior, which is due to the dominance of self-gravity over the tidal potential in the unbound tail. Right: The ratio of the transverse sound-crossing time, $t_{\rm sc}$, to the debris return time, $t_{\rm ret}$, as a function of $\tau$. The vertical, dashed lines indicate the times at which the fluid elements reach their apocenters. The time at which any of the curves exceeds unity (shown by the horizontal, solid black line) is the time after which the fluid element is causally disconnected across its width, which in all cases coincides nearly with when the fluid element reaches apocenter.}
    \label{fig:tsc}
\end{figure*}

The left panel of Figure \ref{fig:tsc} gives the solution to Equation \eqref{Hfin} for $\mu = 4$, which is the expected value for a low-mass star given the ratio of its central to average density and the tidal radius \citep{Coughlin:2022b, Coughlin:2023}, as a function of $\xi_0$ for the $\tau$ shown in the legend. Note that $\xi_0$ ranges from $0.99$ to $1.01$, which is appropriate to a black hole with mass $10^6 M_{\odot}$. The initial behavior (for $\tau \lesssim 4$) is a monotonic increase in $H$ that is independent of $\xi_0$, which is simply a reflection of the fact that $H \simeq e^{\tau/2}$ (independent of $\xi_0$) while each fluid element is on nearly the same, effectively marginally bound orbit. After reaching apocenter and upon returning to the black hole, the tidal field compresses the stream, resulting in a width that approaches $H \simeq 1$ as the material reaches pericenter. Unbound segments of the stream become dominated by self-gravity, which is the origin of the non-monotonic behavior for $\xi_0 > 1$. 

The right panel of this figure shows the ratio of the sound-crossing time to the return time as a function of $\tau$, which can be thought of as the time at which some fictitious perturbation is applied that then ``crosses'' the stream, for $\mu = 4$, where the different curves are for fluid elements specified by the  $\xi_0$ in the legend. Each curve is terminated when $\xi = 1$, as at this point the finite angular momentum of the debris becomes important and the width approaches a non-zero minimum. The vertical, dashed lines delimit the times at which the fluid elements reach their apocenters. This figure demonstrates that even though the fluid elements are tidally compressed as they return to pericenter (as shown by the left panel), the intense shear along the stream results in a monotonically increasing ratio $t_{\rm sc}/t_{\rm ret}$ that equals unity just after apocenter. The stream is therefore causally disconnected at all radii for which the radial velocity is negative and for which the fluid elements are post-apocenter. 

\begin{figure}
    \includegraphics[width=0.48\textwidth]{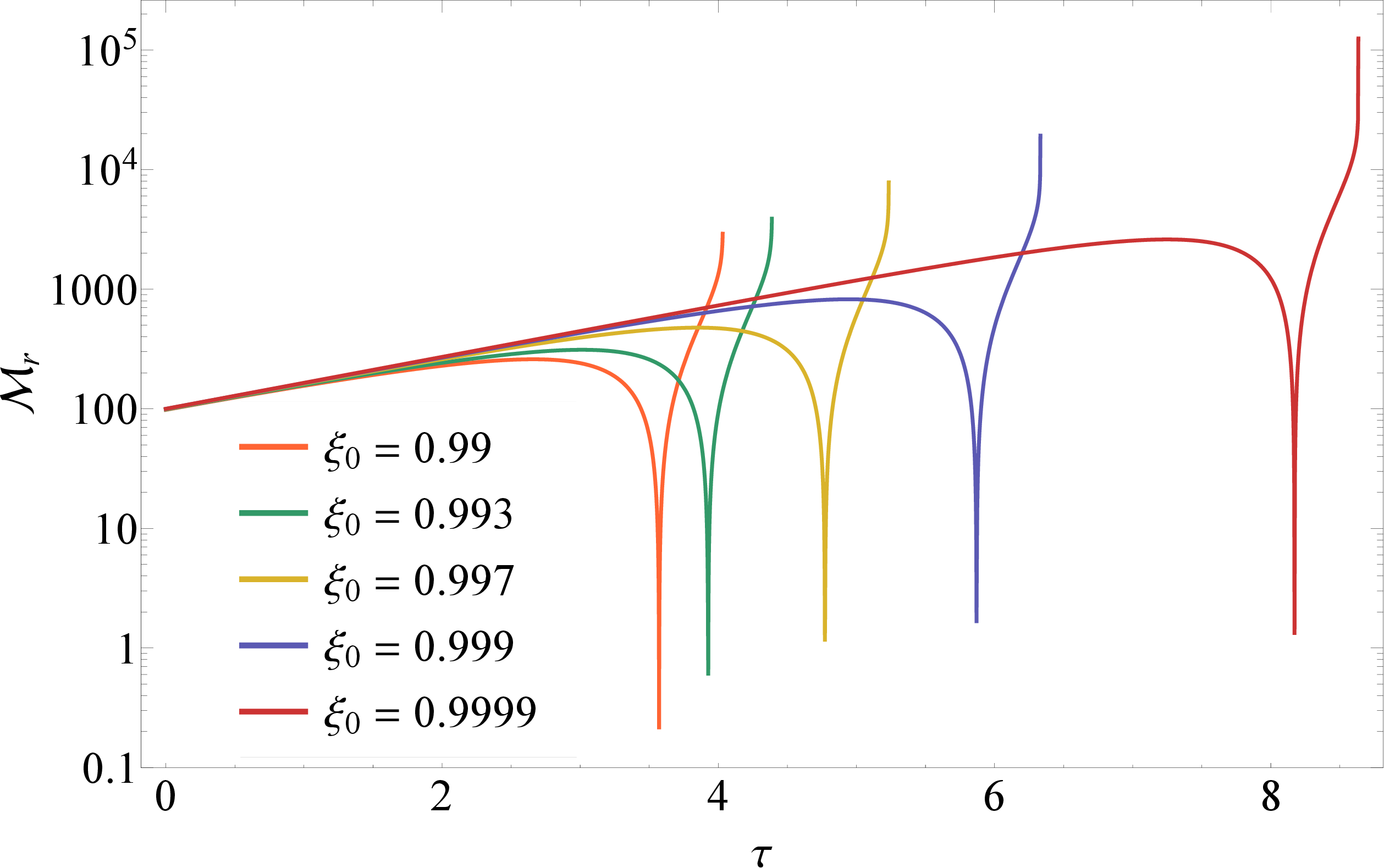}
     \caption{The Mach number associated with the radial velocity for the fluid elements with initial positions given in the legend; here we let $M_{\bullet}/M_{\star} = 10^6$, and otherwise the parameters are the same as those used in Figure \ref{fig:tsc}. The Mach number rises as the sound speed declines, reaches a relative maximum before equaling zero at apocenter, and rapidly rises again to become highly supersonic near pericenter. The curves are stopped when $\xi = 1$, i.e., when the fluid elements return to pericenter.}
    \label{fig:mach}
\end{figure}

We can also determine the radial Mach number (i.e., the Mach number associated with the velocity along the stream): from Equation \eqref{ssdef} we have
\begin{equation}
\begin{split}
    \mathscr{M}_{\rm r} &= \frac{|v|}{c_{\rm s}} = \sqrt{\frac{2GM_{\bullet}}{R_{\rm i}H_{\rm i}^2}}\frac{1}{\sqrt{\gamma\pi G \rho_{\rm c}}}H^{2/3}e^{-\tau/6}|f(\xi)|\left(\frac{\partial\xi}{\partial\xi_0}\right)^{1/3} \\
    & \simeq \left(\frac{M_{\bullet}}{M_{\star}}\right)^{1/3}H^{2/3}e^{-\tau/6}|f(\xi)|\left(\frac{\partial\xi}{\partial\xi_0}\right)^{1/3}, \label{Machrad}
\end{split}    
\end{equation}
where in the last line we let $R_{\rm i} = r_{\rm t}$, $H_{\rm i} = R_{\star}$, and dropped order-unity factors. Figure \ref{fig:mach} shows the Mach number as a function of $\tau$ for the same fluid elements as in the right panel of Figure \ref{fig:tsc}; here we let $M_{\bullet}/M_{\star} = 10^6$ and otherwise the parameters are the same as those chosen for Figure \ref{fig:tsc}. For each fluid element the Mach number starts at a value of 100, which is expected from the fact that the tidal radius is only $\sim 50$ gravitational radii for a sun-like star and a $10^6 M_{\odot}$ black hole (and hence the star's speed at pericenter is $\sim 0.2 c$, compared to its escape speed of $\sim 0.002 c$). Each curve then reaches a relative maximum before dropping to zero at apocenter, becoming hypersonic again as pericenter is reached.

\begin{figure*}
    \includegraphics[width=0.495\textwidth]{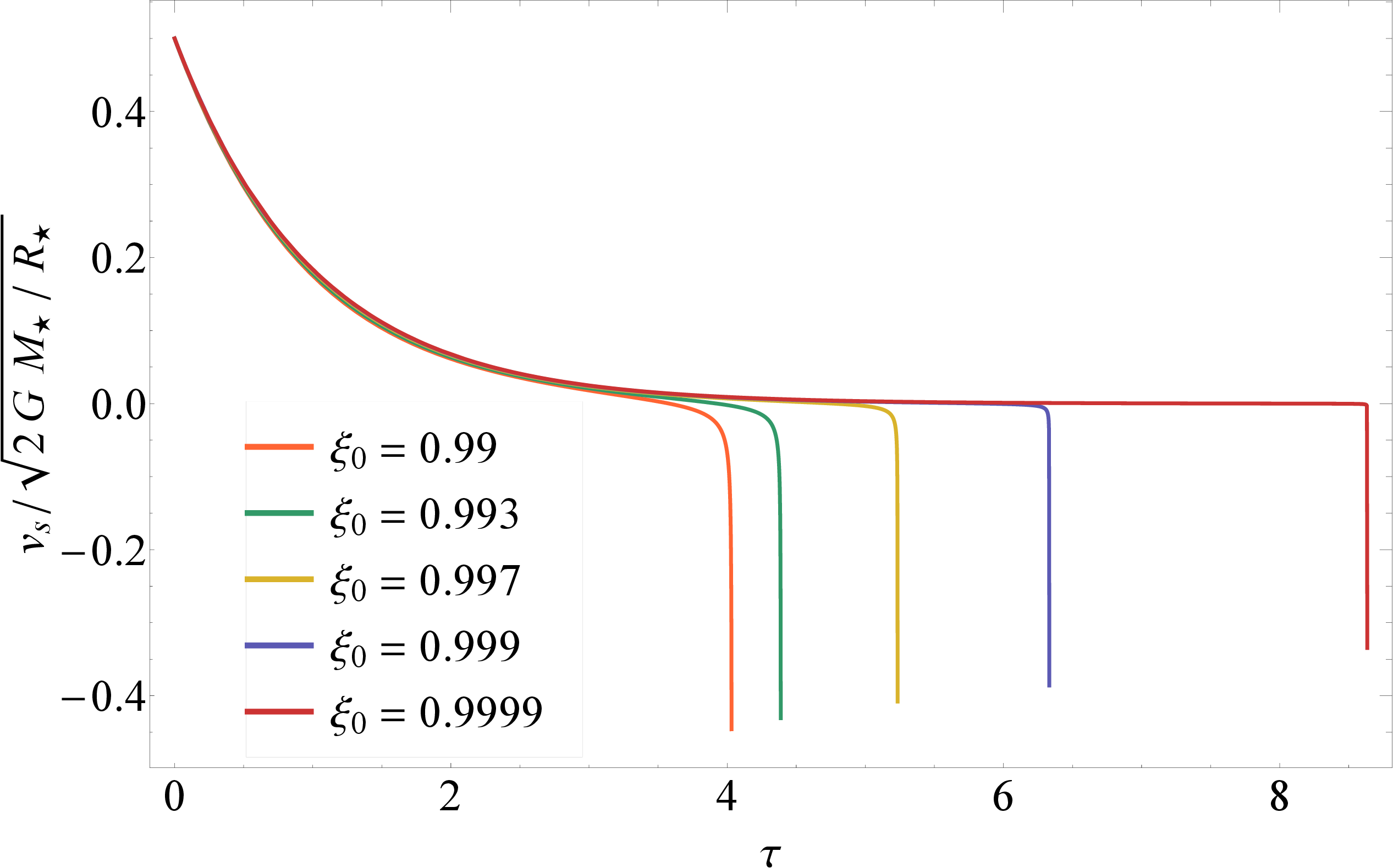}
    \includegraphics[width=0.495\textwidth]{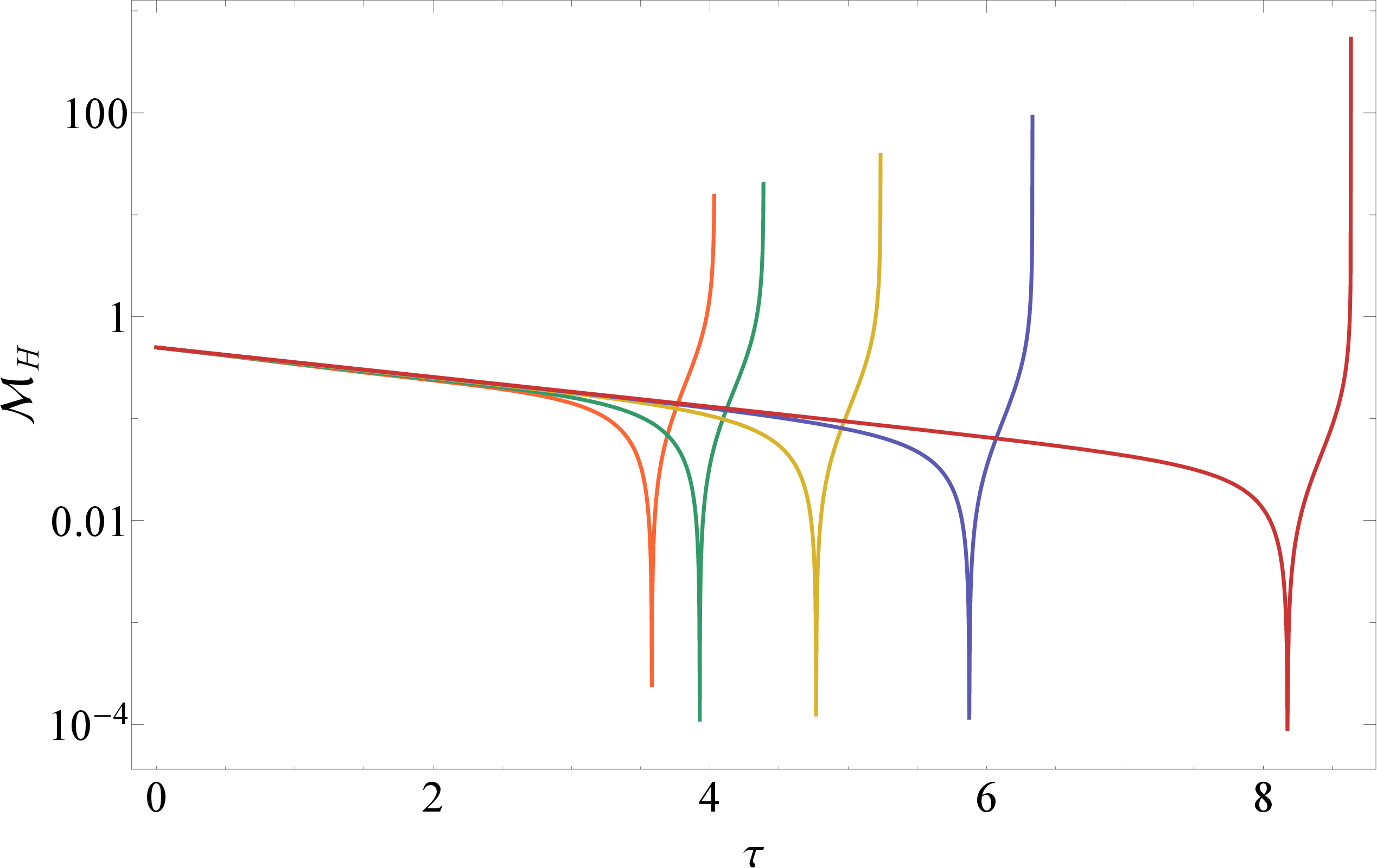}
    \caption{The cylindrical-radial velocity relative to the original stellar escape speed (left) and the cylindrical-radial Mach number (right) as functions of dimensionless time for the same $\xi_0$ as in Figure \ref{fig:mach}. This shows that the velocity out of the plane never exceeds $\sim$ the stellar escape speed and that the Mach number can be large in an absolute sense, but it is orders of magnitude smaller than that associated with the radial velocity (as shown in Figure \ref{fig:mach}).}
    \label{fig:vs}
\end{figure*}

Finally, the cylindrical-radial component of the fluid velocity, i.e., in the $s$-direction, is
\begin{equation}
\begin{split}
    v_{\rm s} &= \frac{\partial H}{\partial t}s_0 = \frac{\sqrt{2GM_{\bullet}}}{R_{\rm i}^{3/2}}e^{-3\tau/2}\frac{\partial H}{\partial \tau}s_0 \\
    &\simeq \sqrt{\frac{2GM_{\star}}{R_{\star}}}e^{-3\tau/2}\frac{\partial H}{\partial \tau},
\end{split}
\end{equation}
where we used $R_{\rm i} = r_{\rm t}$ in the final line and set $s_0 = R_{\star}$, corresponding to a fluid element at the initial stellar surface, while the Mach number associated with this component of the velocity is
\begin{equation}
    \mathscr{M}_{\rm H} = \frac{|v_{\rm s}|}{c_{\rm s}} \simeq e^{-7\tau/6}\left(\frac{\partial \xi}{\partial \xi_0}\right)^{1/3}\left|\frac{\partial H}{\partial \tau}\right|. \label{Machs}
\end{equation}
As for Equation \eqref{Machrad} above, the $\simeq$ arises from dropping order-unity factors (which, because of our simplifying assumption of an initially uniform density along the stream, are inaccurate compared to the realistic situation anyway). Figure \ref{fig:vs} shows the cylindrical-radial velocity relative to the stellar escape speed in the left panel, and the Mach number associated with this component of the velocity in the right, for the same set of $\xi_0$ as in Figures \ref{fig:tsc} and \ref{fig:mach}. We see that the speed in the transverse direction never exceeds $\sim$ the stellar escape speed (which is trivially and immediately satisfied if one ignores the thermal energy of the gas, as the motion is then ballistic in a conservative potential), and the Mach number -- while large in an absolute sense -- is correspondingly much smaller than that associated with the radial velocity.

We assumed that the density is independent of position along the stream (i.e., of $\xi_0$), which was done for simplicity -- including additional variation from, e.g., the initial density profile of the star (e.g., via the methodology in \citealt{Lodato:2009}) is trivial if the pressure gradient and self-gravitational field are not included self-consistently in the dynamics. A more realistic density profile, such that the sound speed approaches zero at the radial extremities of the stream, would lead to the ratio $t_{\rm sc}/t_{\rm ret}$ falling below unity at earlier times and being considerably larger upon returning to pericenter than is indicated for $\xi_0 = 0.99$ in the right panel of Figure \ref{fig:tsc}; similarly, the Mach number of the most-bound fluid elements for a more realistic TDE debris stream would be much higher than the (already-large) values shown in Figure \ref{fig:mach}, and this can be seen directly from Figure \ref{fig:mach-phantom} below. These features have strong bearing on the ability of numerical methods to resolve the stream dynamics as it returns to pericenter, which we now discuss. 

\section{The origin of the anomalous dissipation}
\label{sec:Origin}
Figure \ref{fig:mach} highlights that the Mach number associated with the in-plane motion of the returning debris is at least $\gtrsim 10^3$. However, supersonic motion is obviously a necessary but not sufficient condition for the formation of shockwaves and hydrodynamic dissipation, as (e.g.)~the initial star suffers no such dissipation despite having a Mach number $\gtrsim 100$ as it passes through pericenter. The additional requirement is that the \emph{relative} motion of fluid elements be supersonic, which is (also obviously) absent when the entire star moves nearly with the same velocity.

Because of its hypersonic nature, the returning debris is moving (for all intents and purposes) ballistically, and hence the only shocks that form will do so as a result of caustics. One such caustic is in the out-of-plane direction (i.e., in the direction parallel to the orbital angular momentum vector of the original star), which has been the subject of extensive discussion and debate (e.g., \citealt{Kochanek:1994, Guillochon:2014, Bonnerot:2022, Andalman:2025}). 
Figure \ref{fig:vs} illustrates that the out-of-plane motion never exceeds the stellar escape speed, but the Mach number can still be appreciable, and hence there can be shocks\footnote{Note that this is not a given, nor is it addressable with the model developed in Section \ref{sec:analytic}, because a homologous velocity profile (which we assumed) yields a one-to-one mapping between the initial and current Lagrangian positions of each fluid element and precludes shock formation. Indeed, shocks formed from the gravitational compression of gas are typically much weaker than simple Mach-number-based estimates imply, as the increasing density leads to adiabatic heating and the increasing sound speed works to resist the formation of a shock \citep[cf.][]{Norman:2021,Coughlin:2022}.} that form as the stream is compressed (though radiative recombination will raise the temperature relative to the adiabatic one assumed here, which will increase the sound speed and reduce the effective Mach number; see \citealt{Andalman:2025}). Nevertheless, because the speed itself is 2-3 orders of magnitude smaller than the in-plane speed and dissipation will at most yield a thermal pressure comparable to the vertical ram pressure, its impact on the in-plane motion is ignorable; this conclusion agrees with the results of, e.g., \citet{Guillochon:2014, Bonnerot:2022, Andalman:2025}. 

There is no analogous caustic formation within the plane (this is clear kinematically and as we show imminently, but see also \citealt{Kubli:2025}, whose $10^{10}$-particle simulation shows that there is little-to-no spraying or dissipation at pericenter), and hence the only site of (weak) dissipation \emph{should} be that associated with the vertical motion, implying that a dynamically cold stream should emerge from the pericenter passage and impact the incoming stream at effectively a single point. However, it is near pericenter that the nature of the in-plane velocity field changes qualitatively and in a way that is problematic for standard numerical methods. Specifically, as the stream is returning to but has not yet reached pericenter, the velocity along the stream is hypersonic and diverging implying that numerically induced viscous effects (i.e., numerical dissipation) are negligible: the artificial viscosity in an SPH simulation is not applied to receding particle pairs \citep[e.g.,][for the specific implementation in {\sc phantom}; see also the next subsection]{Price:2018}, while the solution to the Riemann problem at the interface between two cells in a Godunov-based scheme is two rarefaction fans. Upon reaching pericenter, the velocity profile  along the stream changes from diverging to converging, as fluid elements start to decelerate and undergo their second orbit about the black hole. In the limit of a parabolic orbit (for simplicity and because it is highly accurate for typical massive black hole-star encounters), the component of the velocity that is perpendicular to the direction of pericenter and within the orbital plane of the original star is
\begin{equation}
    v(y) = \sqrt{\frac{2GM}{r_{\rm p}}}\left(1-\frac{1}{4}\left(\frac{y}{r_{\rm p}}\right)^2\right)+\mathcal{O}\left[\left(y/r_{\rm p}\right)^{4}\right], \label{vyeq}
\end{equation}
where $y$ is in the direction parallel to $v$ and $r_{\rm p}\simeq r_{\rm t}$ is the pericenter distance. Therefore and in the limit of infinite resolution, the kinematic in-plane velocity profile does not lead to the formation of a shock in the immediate vicinity of pericenter at $y = 0$. However, with finite resolution and particularly if the particle smoothing length is comparable in size to the pericenter distance -- which will typically be the case for the leading edge of the stream that originated near the stellar surface -- a large convergence in the velocity field (i.e., across neighboring particles) can be spuriously produced, leading to correspondingly large viscous forces on the particles with precise magnitudes that vary stochastically with time. 
Analogously, with an insufficiently small cell size near pericenter, the finite difference in velocities of neighboring cells will lead to strong-shock solutions to the Riemann problem because of the highly supersonic nature of the flow, leading to non-physical dissipation and heating of the gas.

We can more directly and analytically investigate the nature of the velocity field of the returning debris, but the model in the preceding section adopts a purely radial morphology of the gas, i.e., it is one-dimensional both in terms of the spatial and velocity distributions of the fluid and therefore cannot be used self-consistently to this end. However, \citet{Coughlin:2023} and \citet{Coughlin:2025} formulated a model that is still one-dimensional from the standpoint that there is only a single direction that is ``along the stream,'' but is two-dimensional in that this direction is curvilinear and parameterized by the initial positions of fluid elements under the frozen-in prescription, and it therefore captures the finite angular momentum of the material (but it excludes pressure gradient and self-gravitational forces in the transverse direction). Specifically, the debris satisfies the usual frozen-in condition initially (e.g., \citealt{Lodato:2009}), but the density profile is collapsed into a one-dimensional line that extends purely in the direction of pericenter; see \citet{Coughlin:2025} for more details. From this model, the divergence of the velocity field along the stream is
\begin{equation}
    \nabla\cdot\mathbf{v} = \frac{\frac{\partial X}{\partial X_0}\frac{\partial^2X}{\partial t\partial X_0}+\frac{\partial Y}{\partial X_0}\frac{\partial^2 Y}{\partial t\partial X_0}}{\left(\frac{\partial X}{\partial X_0}\right)^2+\left(\frac{\partial Y}{\partial X_0}\right)^2}. \label{deldotv}
\end{equation}
Here $X_0$ is the initial Lagrangian position of the fluid element with current Lagrangian position $\{X(X_0,t),Y(X_0,t)\}$, which satisfy Equations (19) and (20) from \citet{Coughlin:2025}, $\partial/\partial t$ is the Lagrangian time derivative (i.e., for a given $X_0$), and this expression (i.e., Equation \ref{deldotv}) arises from taking the Lagrangian time derivative of Equation (23) in \citet{Coughlin:2025}. 

\begin{figure*}
    \includegraphics[width=0.495\textwidth]{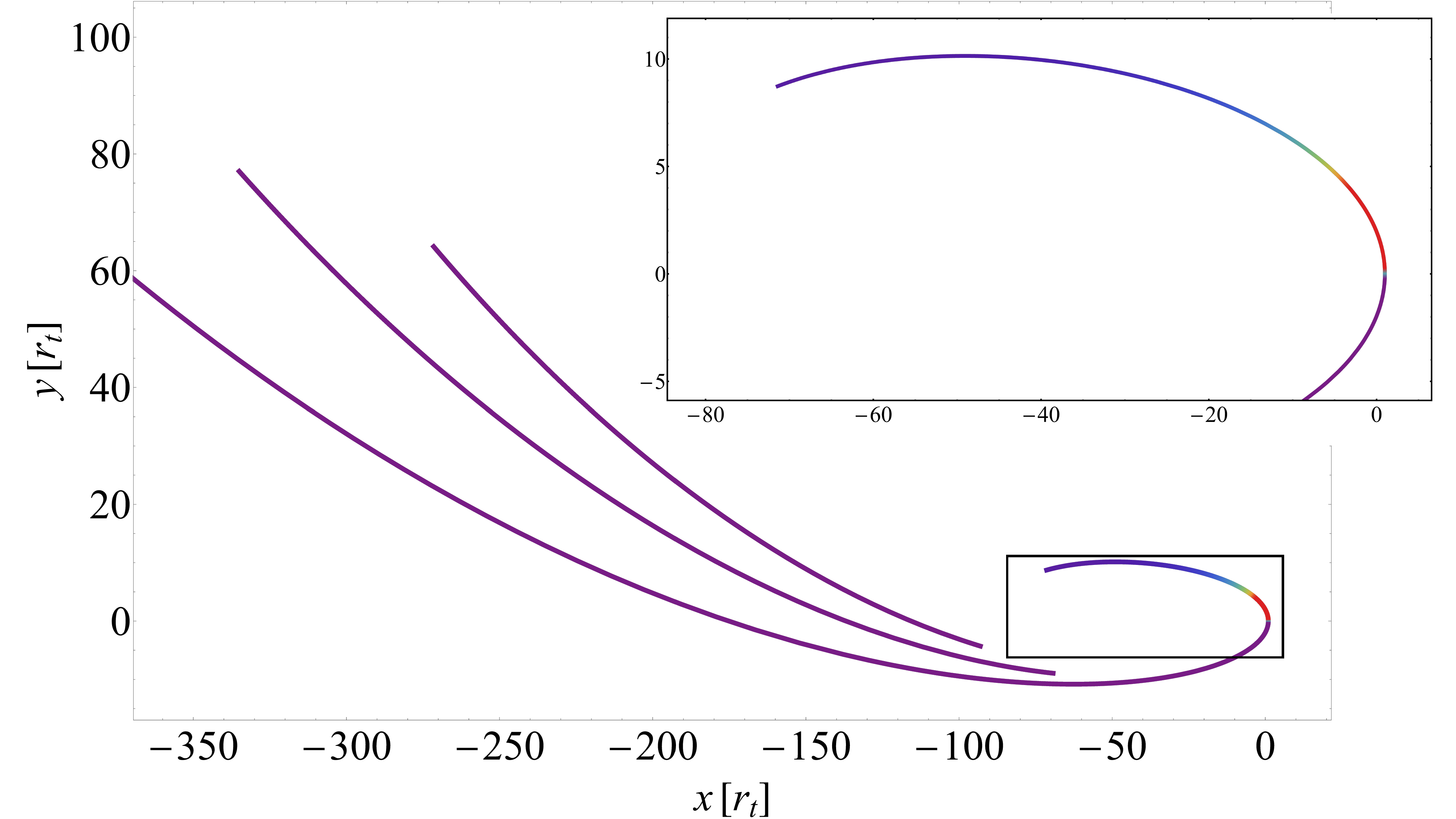}
    \includegraphics[width=0.495\textwidth]{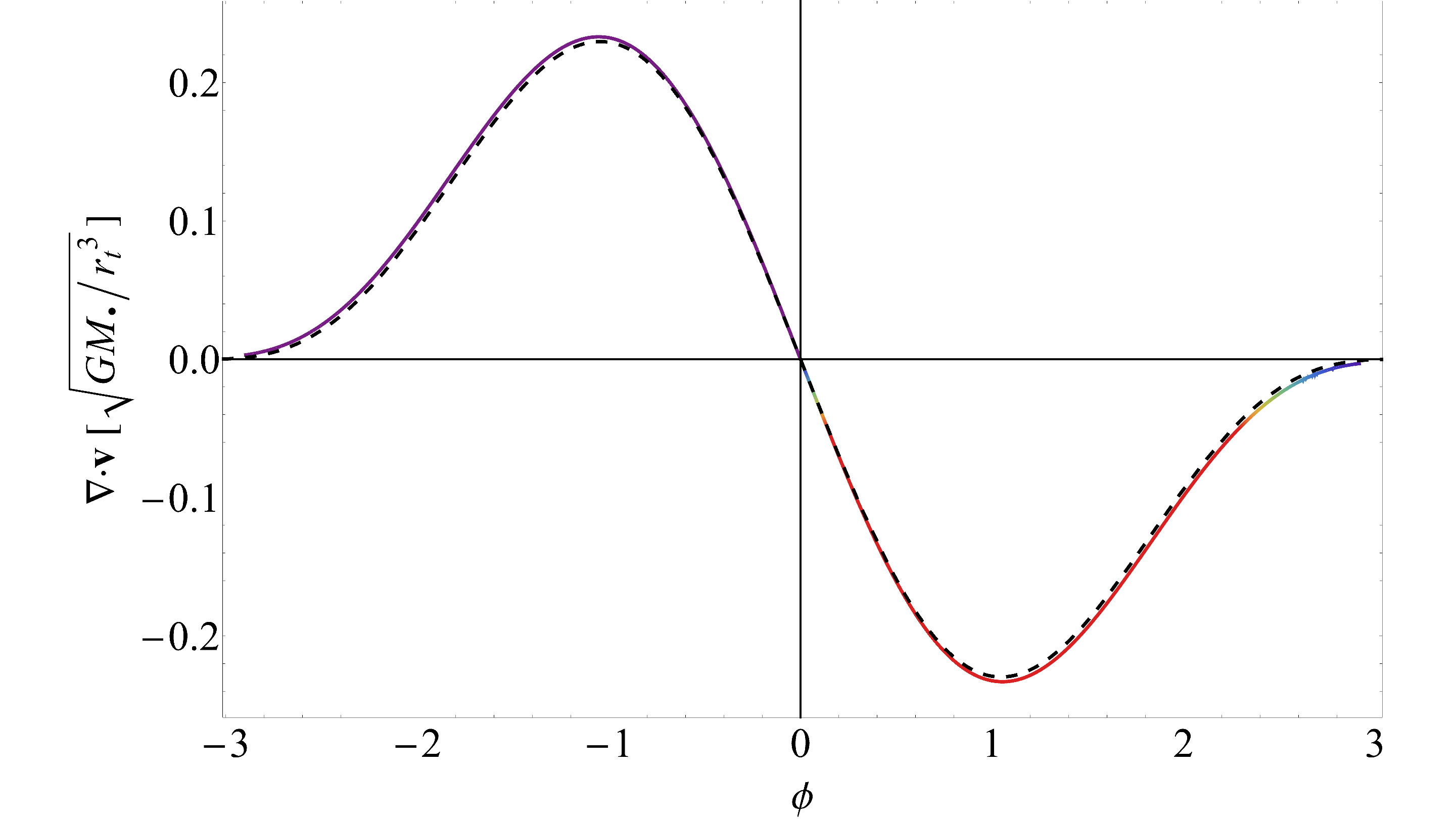}
    \caption{Left: the stream of debris predicted from the one-dimensional model in \protect\cite{Coughlin:2025} at three different times post-disruption, following the disruption of a solar-like star by a $10^6 M_{\odot}$ black hole. Each curve is color-coded in a way that scales with the local value of $\nabla\cdot\mathbf{v}$, such that purple coincides with regions of diverging flow ($\nabla\cdot\mathbf{v} > 0$) while blue-to-red have converging flow ($\nabla\cdot\mathbf{v} < 0$). The inset provides a zoom-in on the fluid elements that have already returned to pericenter and highlights the fact that they are in regions of converging flow. Right: the divergence of the fluid elements, in units of the inverse of the dynamical time at the tidal radius, as a function of the angle $\phi$ that is measured from the direction of pericenter (a.k.a.~the true anomaly). The coloration corresponds to that in the left panel, highlighting that there is a region of strong convergence as the fluid elements pass through pericenter, and the black-dashed curve gives the analytical prediction for a zero-energy Keplerian orbit (Equation \ref{deldotv2}). }
    \label{fig:streams_w_ddotv}
\end{figure*}

Figure \ref{fig:streams_w_ddotv} shows the predicted (from this model) distribution of the debris at three different times post-disruption, two prior to the return of the most-bound debris and one after, following the disruption of a solar-like star by a $10^6 M_{\odot}$ black hole. The coloration of the curves scales with the local value of $\nabla\cdot\mathbf{v}$ (given by Equation~\ref{deldotv}), such that purple coincides with regions of diverging flow ($\nabla\cdot\mathbf{v} > 0$), while blue, green, yellow, orange, and red represent regions of converging ($\nabla\cdot\mathbf{v} < 0)$ flow and in order of increasing magnitude of convergence. The inset provides a zoom-in on the fluid that has returned to pericenter. The right panel of this figure shows the local divergence of the fluid velocity, measured in units of the inverse of the dynamical time at the tidal radius, as a function of the angle of the orbit that is measured with respect to the direction of pericenter, $\phi$, i.e., the true anomaly. The coloration in this panel reflects that in the left panel: the fluid elements experience a strong convergence as they pass back through pericenter, resulting in the compression of the gas. However, the magnitude of $\nabla\cdot\mathbf{v}$ remains finite everywhere, indicating that there is no caustic and hence no shock formation within the flow -- the ordering of the fluid elements is preserved and there is no interpenetration of the gas. The dashed curve in this panel is the prediction that follows from zero-energy Keplerian orbits, being
\begin{equation}
\begin{split}
    \nabla\cdot\mathbf{v} &= -\frac{1}{4}\sqrt{\frac{GM_{\bullet}}{2 r_{\rm t}^3}}\sin\phi\left(1+\cos\phi\right) \\
    &\simeq -\sqrt{\frac{GM_{\bullet}}{8 r_{\rm t}^3}}\frac{y}{r_{\rm t}}. \label{deldotv2}
\end{split}
\end{equation}
Note that this is a factor of 2 smaller compared to what is obtained by taking $\partial v/\partial y$ with $v$ given by Equation \eqref{vyeq}, because there is an $x$-component of the velocity that increases as the gas recedes to larger radii, i.e., Equation \eqref{deldotv} correctly accounts for both the magnitude and the direction of the velocity as one moves along the curvilinear axis represented by the stream. The small differences between Equation \eqref{deldotv} (the exact solution) and \eqref{deldotv2} are from the small but non-zero binding energies of the fluid elements.

We propose that it is the qualitative change in the stream velocity profile through pericenter (shown by Equations~\ref{deldotv} \& \ref{deldotv2} and Fig.~\ref{fig:streams_w_ddotv}), alongside the under-resolved nature of the simulations and the very weakly bound nature of the debris -- such that the work done by the spurious viscous forces on the fluid elements is large compared to their binding energies -- that results in the ``fanning'' and heating of the material as it returns to pericenter. Aside from increasing the particle number, either locally in the case of \citet{Ayal:2000} or globally as done in \citet{Kubli:2025}, to test this hypothesis one can simply reduce the magnitude of the artificial viscosity employed in the SPH algorithm and investigate the differences that arise. We perform this exercise in the next subsection. 

\subsection{Numerical simulations}
\label{sec:simulations}
We test our hypothesis by running and analyzing a set of simulations with the {\sc phantom} SPH code. To model the relativistic apsidal precession of the debris we employ the Einstein potential\footnote{The details of the self-intersection obviously depend on the choice and accuracy of the pseudo-Newtonian potential (e.g., \citealt{Tejeda:2013}), but these details are not the point of the present investigation, and the Einstein potential suffices for our purposes.} for the gravitational field of the black hole, given by \citep{Nelson:2000}
\begin{equation}
    \Phi(r) = -\frac{GM}{r}\left(1+\frac{3R_{\rm g}}{r}\right)\,,
\end{equation}
where $R_{\rm g}=GM_{\bullet}/c^2$ is the gravitational radius. We also include an accretion radius of $4R_{\rm g}$, inside of which particles are removed from the simulation. The star is taken to be an $n=3/2$ polytrope with a solar mass and solar radius, and placed on a parabolic orbit with pericentre equal to the tidal radius $r_{\rm t} = (M/M_\star)^{1/3}R_\star$. The equation of state for the gas is $P = K\rho^\gamma$ where $\gamma = 5/3$ is a fixed constant. The entropy function $K$ is fixed in simulations that exclude heating from the artificial viscosity terms (i.e., the entropy equation is trivially $\dot{K} = 0$), while for simulations that include heating from the artificial viscosity, $K$ is evolved to include this effect \citep[see][for details]{Price:2018}. We perform the simulations at $10^5$, $10^6$ and $10^7$ particles to simultaneously investigate the effects of resolution.

Since our main interest here is to understand the impact of the artificial viscosity on the simulation results, we briefly summarize its standard implementation in the SPH algorithm. Specifically, the artificial viscosity is composed of two terms: a linear (quadratic) term that is linear (quadratic) in the fluid velocity. For each particle, $a$, its pressure, $P_a$, is augmented in the equation of motion (i.e., the momentum equation) by an amount $q_{ab}^a$ (Equation~34 in \citealt{Price:2018}), where $q_{ab}^{a}$ is given by
\begin{equation}
    q_{ab}^a =  \begin{cases}
                    -\frac{1}{2}\alpha_a^{\rm AV}\rho_a v_{{\rm sig},a}\mathbf{v}_{ab}\cdot \hat{\mathbf{r}}_{ab} & \mathbf{v}_{ab}\cdot \hat{\mathbf{r}}_{ab} < 0\\
                    0 & {\rm otherwise}
                \end{cases}. \label{qab}
\end{equation}
Here the signal velocity $v_{{\rm sig},a} = c_{{\rm s},a} + \beta^{\rm AV}|\mathbf{v}_{ab}\cdot \hat{\mathbf{r}}_{ab}|$, $\mathbf{v}_{ab} = \mathbf{v}_a-\mathbf{v}_b$, and $\mathbf{\hat{r}}_{ab} = (\mathbf{r}_a-\mathbf{r}_b)/|\mathbf{r}_a-\mathbf{r}_b|$. As mentioned above, there is only a viscous contribution to the equation of motion when the relative velocity between two particles is negative, i.e., when the flow is compressive and $\mathbf{v}_{ab}\cdot\hat{\mathbf{r}}_{ab} < 0$. Also note that Equation \eqref{qab} represents a slight adaptation to Equations (40) and (41) in \cite{Price:2018} to include the Cullen-Dehnen switch (used to reduce the value of $\alpha_a^{\rm AV}$ away from regions not conducive to shock formation; see Equation \ref{cd} below and the discussion thereof) in the quadratic term as well as the linear term \citep{Cullen:2010,Chen:2025}.\footnote{Although, in this case, whether or not the switch is applied to the quadratic term makes little difference in practice as $\alpha^{\rm AV} \simeq 1$ for all particles that have returned close enough to the black hole.}
From this we can see that the linear term scales as $\sim c_sv$ and the quadratic term scales as $\sim v^2$ \citep[cf.~e.g.][]{Monaghan:1997,Lodato:2010}, with the proportionality constant being $\alpha^{\rm AV}$ for the linear term and the product of $\alpha^{\rm AV}$ and $\beta^{\rm AV}$ for the quadratic term. As shown in Section~\ref{sec:analytic} (see also Figure \ref{fig:mach-phantom} below), the Mach number of the debris stream, $\mathscr{M} = v/c_{\rm s} \gg 1$, and therefore we can expect that where artificial viscosity is applied it is the quadratic term that dominates. However, when heating due to artificial viscosity is included, any strong shocks in the flow will raise the sound speed such that $c_{\rm s}\sim v$, at which point the linear and quadratic viscosities have comparable magnitudes.

Given these considerations, we perform an SPH simulation with $\alpha^{\rm AV} = \beta^{\rm AV} = 0$ (no artificial viscosity), and five additional simulations in which the value of $\beta^{\rm AV}$ is varied from zero to the standard value of 2. In the additional five simulations the value of $\alpha^{\rm AV} \in [0,1]$ varies for each particle and is set via the Cullen-Dehnen switch (with the specific implementation as outlined in Section 2.2.9 of \citealt{Price:2018}). This leads to the $\alpha^{\rm AV}$ for each particle being set to a value of $\alpha_{\rm loc}^{\rm AV}$, given by 
\begin{equation}\label{cd}
\alpha_{\rm loc}^{\rm AV} = \min\left\{\frac{10h^2}{c_s^2}\chi\max\left[-\frac{\rm d}{{\rm d}t}\left(\nabla\cdot\mathbf{v}\right),0\right],\alpha_{\rm max}^{\rm AV}\right\},
\end{equation}
where $\chi$ is the \cite{Balsara:1995} limiter given by
\begin{equation}
    \chi = \frac{|\nabla\cdot\mathbf{v}|^2}{|\nabla\cdot\mathbf{v}|^2 + |\nabla\times\mathbf{v}|^2} \label{balsara}
\end{equation}
and we take $\alpha_{\rm max}^{\rm AV}=1$, or, if the value of $\alpha^{\rm AV}$ for the particle from the previous timestep is greater than $\alpha_{\rm loc}^{\rm AV}$, then $\alpha^{\rm AV}$ is evolved (decayed) as $\dot{\alpha}^{\rm AV} = -(\alpha^{\rm AV}-\alpha_{\rm loc}^{\rm AV})/\tau$ where $\tau = 10h/c_{\rm s}$.

\begin{figure}
    \includegraphics[width=0.475\textwidth]{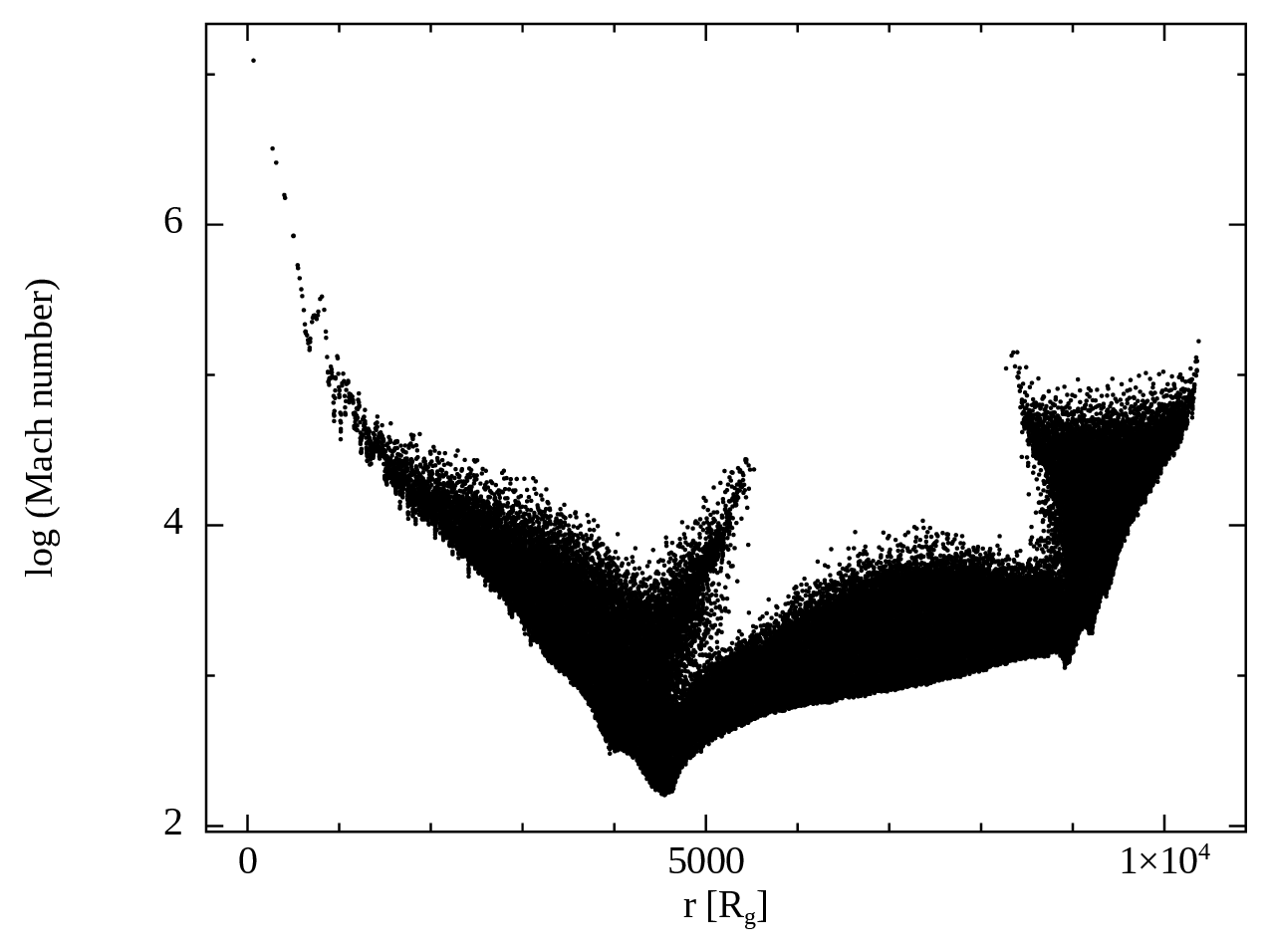}
    \caption{The Mach number of the fluid elements following the disruption of a solar-type, $5/3$-polytrope by a $10^6 M_{\odot}$ supermassive black hole, at a time when the most-bound fluid element is just returning to pericenter (roughly 18.5 days since the star was at its original point of closest approach to the black hole); this fluid element is in the top-left corner of the plot and has a Mach number of $\sim 10^{7}$. The particle number is $10^6$. The horizontal axis gives the distance of the fluid elements from the black hole in units of gravitational radii, and here the artificial viscosity parameters are $\alpha^{\rm AV} = \beta^{\rm AV} = 0$. In agreement with the estimates in Section \ref{sec:analytic}, this illustrates that the flow is hypersonic upon returning to pericenter.}
    \label{fig:mach-phantom}
\end{figure}

Figure \ref{fig:mach-phantom} illustrates the fluid Mach number as a function of distance from the black hole (in units of gravitational radii) from the simulation with $\alpha^{\rm AV} = \beta^{\rm AV} = 0$ and $10^6$ particles. This distribution of particles coincides with the time at which the most-bound fluid element, which is in the top-left corner of the figure with a Mach number of $\sim 10^{7}$, has just returned to pericenter. In agreement with the analytical model in Section \ref{sec:analytic}, this figure demonstrates that the fluid is highly supersonic, especially for the low-density tails of the debris that comprise fluid elements that were initially near the stellar surface at the time of disruption. 

\begin{figure*}
    \centering
    \includegraphics[width=\textwidth]{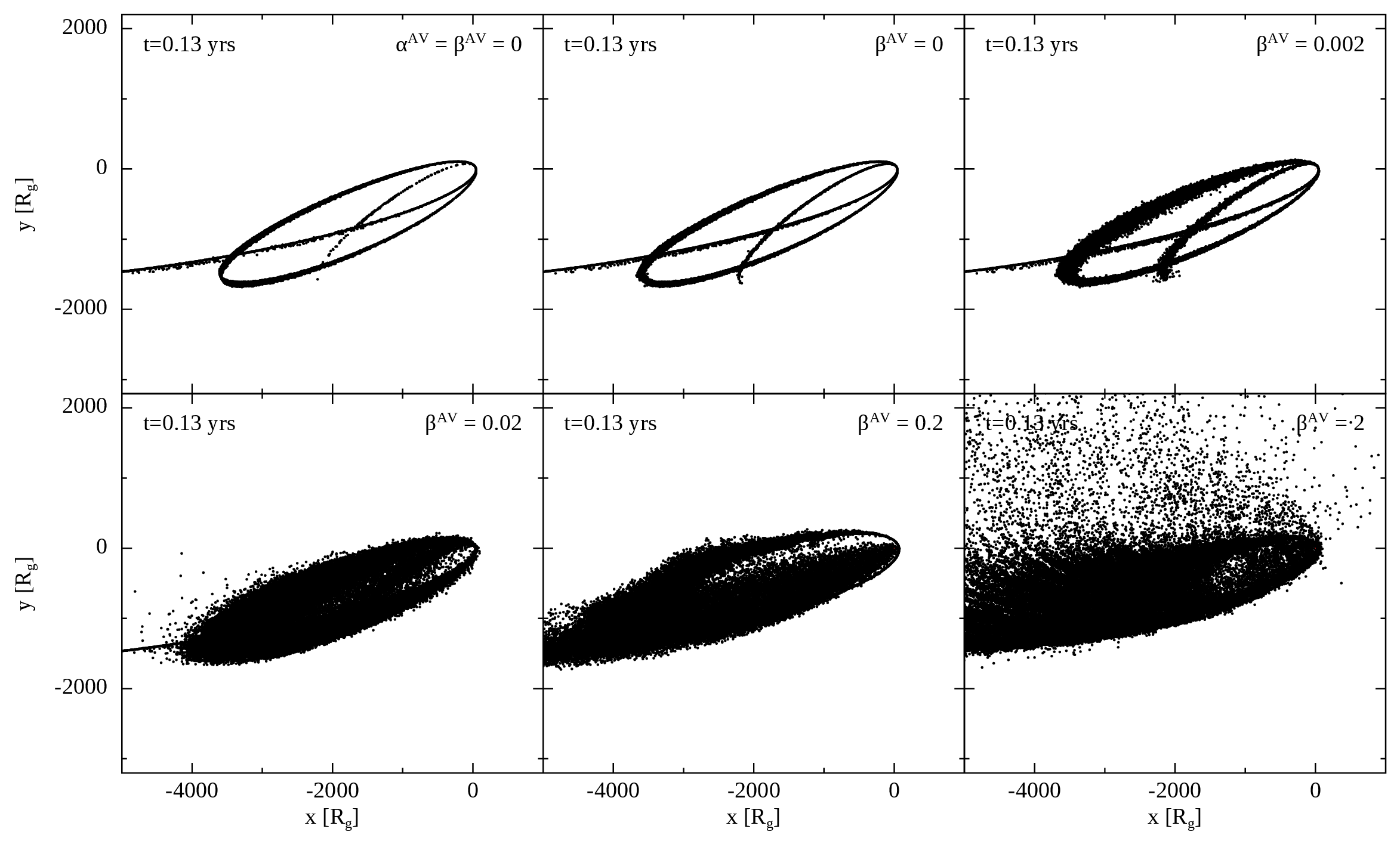}
    \caption{The returning stream structure with varying amounts of quadratic artificial viscosity, denoted by the value of $\beta^{\rm AV}$ given in the panels, for a $5/3$-polytropic star disrupted by a $10^6 M_{\odot}$ black hole. In all simulations shown the star was modeled with $10^6$ SPH particles. In each case the stream enters from the bottom left (coinciding with $\{x,y\} \approx \{-5,000,-1,500\}$) with particles moving toward the black hole, which is located at $(0,0)$. By the time depicted in the plots ($0.13$\,yrs after the initial disruption of the star) the most bound material has had time to do $\approx 2.5$ orbits around the black hole. For each simulation the linear artificial viscosity is included by employing the default Cullen-Dehnen switch, except for the top-left panel where it is manually set to zero. For $\beta^{\rm AV} \lesssim 0.2$ the stream intersection shock is not captured adequately, i.e., for $\beta^{\rm AV} \lesssim 0.002$ the stream largely passes through itself. In this set of simulations the heating of the gas due to the artificial viscosity is not included, and thus the fanning of the debris is entirely due to the communication of the velocity field in the debris by the artificial viscosity terms.} 
    \label{fig:noshock}
\end{figure*}

In Figure \ref{fig:noshock} we show the debris stream structure in simulations with these viscosity coefficients and $10^6$ SPH particles, and for which the heating due to viscosity is not included in the thermal energy of the gas. The figures show a snapshot at a time of 0.13~years after the initial disruption of the star, by which time a significant amount of debris has returned to pericentre (and some of the material has returned twice). In the top-left panel with $\alpha^{\rm AV} = \beta^{\rm AV} = 0$ (i.e., no artificial viscosity is included in the simulation), there is no spraying of the debris as it passes through pericenter. This figure also shows that there is no interaction of the fluid when the streams cross; this unphysical outcome is why artificial viscosity \emph{should} be included in SPH simulations, as it prevents inter-particle penetration (i.e. multi-valued velocity fields) that should correspond physically to the formation of shocks in the flow. As expected, the same structure arises in the simulation with only the linear $\alpha^{\rm AV}$ term (the top-middle panel). A small quadratic term ($\beta^{\rm AV}=0.002$) results in a noticeable broadening of the stream, as shown in the top-right panel, and for larger $\beta^{\rm AV}$ values (the bottom row of this figure) the stream is widened into the standard ``fan'' that is typical of simulations of returning TDE debris streams (and that \citealt{Kubli:2025} showed does not occur once sufficient resolution is reached).

\begin{figure*}
    \centering
    \includegraphics[width=\textwidth]{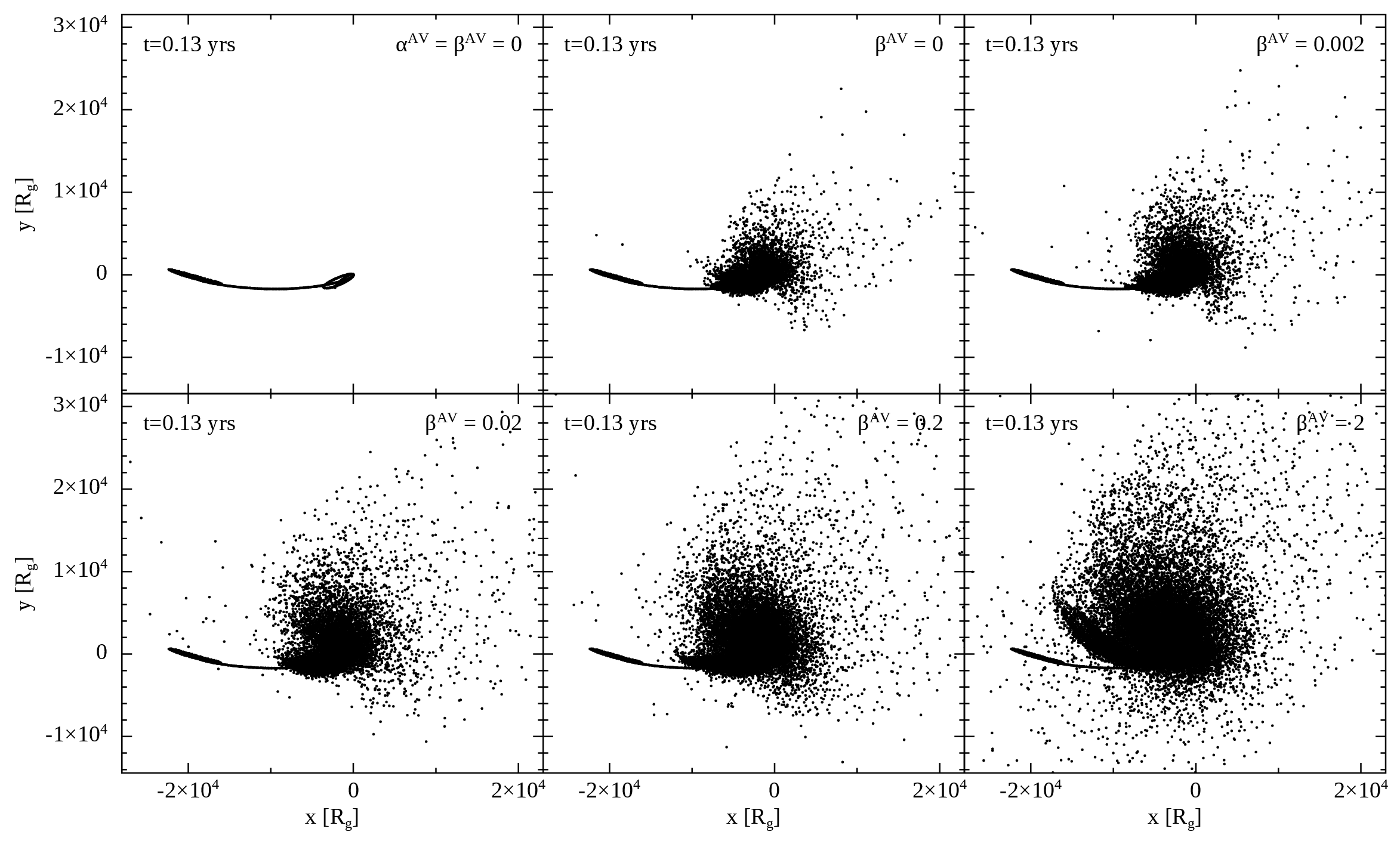}
    \caption{Same as Figure~\ref{fig:noshock}, but here the heating from the artificial viscosity terms is included in the simulations and we have increased the spatial range of each axis to illustrate the spread of particles. With no artificial viscosity (top-left panel) there is no evolution of the stream beyond orbital dynamics. For $\beta^{\rm AV} = 0$ (top-middle panel) there is significant heating and expansion of the stream compared to the analogous panel in Figure \ref{fig:noshock}; this is because the gas is heated by the linear viscosity term until $c_{\rm s} \sim v$, at which point the linear term ($\sim c_{\rm s}v$) is elevated to the same velocity dependence (and level of importance) as the quadratic term ($\sim v^2$). The morphology of the flow depends on the level of artificial viscosity employed, i.e. the $\beta^{\rm AV}$ value, and the size of the emitted cloud monotonically increases with increasing $\beta^{\rm AV}$.}
    \label{fig:shock}
\end{figure*}

Figure \ref{fig:shock} is the same as Figure \ref{fig:noshock}, but in this case the heating from the artificial viscosity terms is included. The no-artificial-viscosity case (and thus no heating) proceeds as before. However, all the other cases display a similar morphology. As discussed above, the linear term is strongly augmented by the heating of the gas, such that it is comparable to the quadratic term. In these cases, increasing the quadratic term serves primarily to increase the size of the ``cloud'' of debris that is generated, though there are also morphological differences that arise from changes to $\beta^{\rm AV}$.

\begin{figure*}
    \centering
    \includegraphics[width=\textwidth]{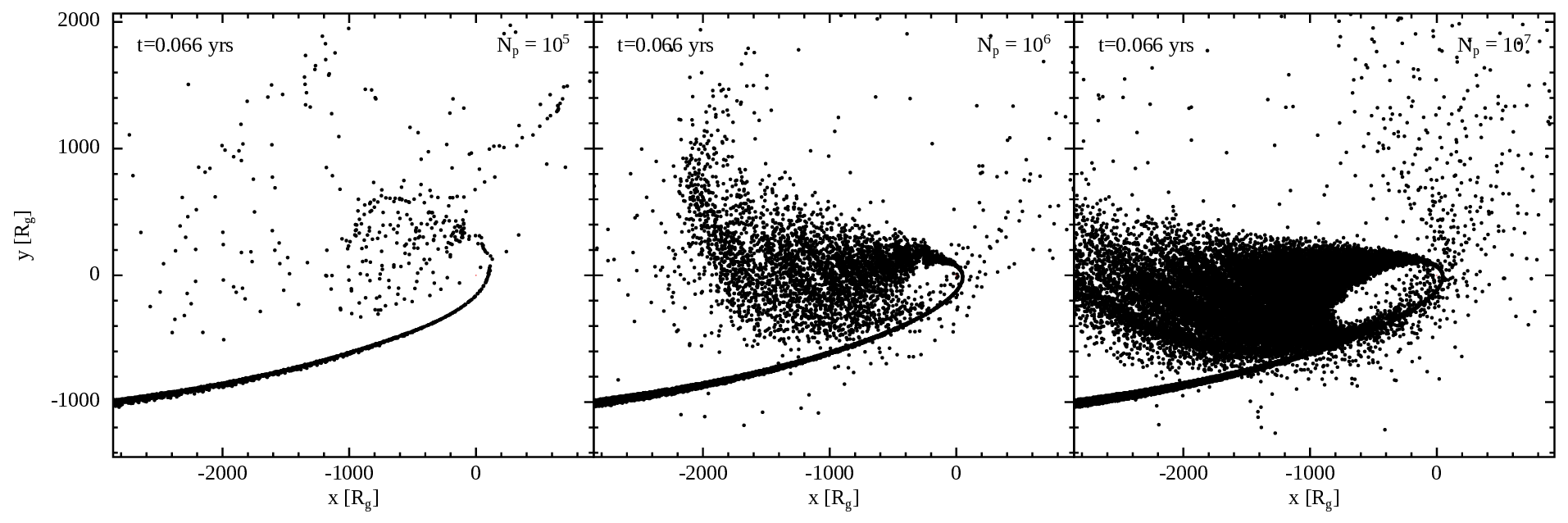}
    \includegraphics[width=\textwidth]{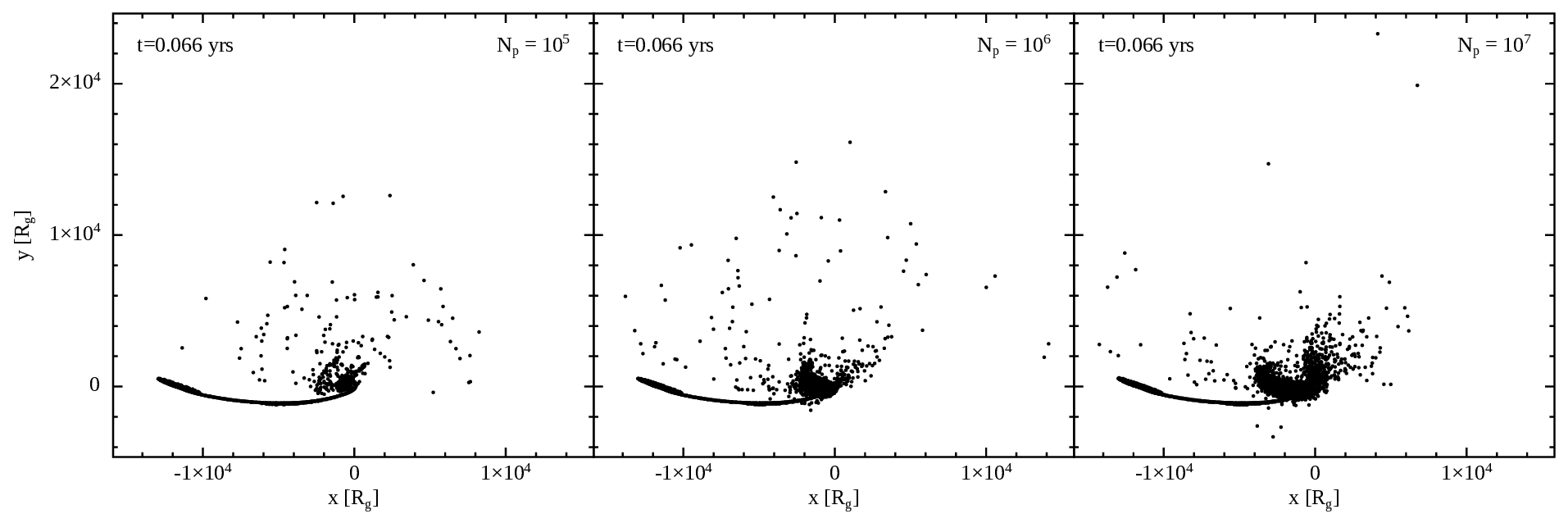}
    \caption{Returning stream structures in simulations with varying particle number, from $10^5$ (left), to $10^6$ (middle) and $10^7$ (right). In each case $\beta^{\rm AV} = 2$ and heating from the artificial viscosity terms is included, i.e. the right hand panels here correspond to the bottom right panel in Figure~\ref{fig:shock}, albeit at an earlier time. The top row of plots demonstrates that TDE simulations of the returning debris with up to $10^6$ particles are not converged: the debris morphology changes significantly with each change in resolution. The bottom row shows a zoomed-out version of the expanding cloud of debris -- its mass, volume and shape change as a function of particle number. The overall trend is that higher resolution leads to less material ejected to large scales, and a less spherical distribution of that debris.} 
    \label{fig:converge}
\end{figure*}

Figure \ref{fig:converge} shows the debris distribution with $\beta^{\rm AV} = 2$, artificial viscosity heating included, and at a time of $0.066\textrm{ yr} \simeq 24\textrm{ days}$ for $10^{5}$ (left), $10^6$ (middle), and $10^{7}$ (right) SPH particles. The bottom row is the same as the top row, but zoomed out to show the entire range of scales to which particles have propagated. This figure demonstrates that there is a large amount of dissipation that occurs near pericenter, i.e., before any self-intersection of the incoming and outgoing debris streams, and in a way that is qualitatively different from one resolution to the next.

\section{Discussion}
\label{sec:discussion}
The simulations described in the preceding section demonstrate that artificial viscosity is entirely responsible for the simultaneous heating and fanning effect that is consistently recovered in SPH simulations (e.g., \citealt{Ayal:2000}), with the morphology of the material changing qualitatively in going from large-to-small viscosity. Indeed, with only $10^5$ particles (and probably fewer) the stream can be kept narrow by removing the SPH artificial viscosity entirely, i.e., by setting $\alpha^{\rm AV} = \beta^{\rm AV} = 0$, while its inclusion requires at least billions of particles to recover accurate simulations \citep{Kubli:2025}. 

Of course, simply setting $\alpha^{\rm AV} = \beta^{\rm AV} = 0$ does not serve to completely remedy the issue, because by simulating a highly supersonic flow with no artificial viscosity we are, in effect, studying the evolution of non-interacting dust in a fixed gravitational potential -- not a fluid. To maintain the fluid nature of the simulation, and thus to adequately describe the true dissipation that occurs as the stream is compressed vertically near pericenter and at the self-intersection point, inter-particle penetration must be prevented via the inclusion of some viscous-type parameterization. Nevertheless, it was shown in Section \ref{sec:analytic} (and in various other investigations, e.g., \citealt{Guillochon:2014, Bonnerot:2022, Andalman:2025}) that the speed of the stream in the out-of-plane direction, and as it is compressed near pericenter, is on the order of the original stellar escape speed. This speed is $\sim 2-3$ orders of magnitude smaller than the in-plane velocity, and hence even if high-Mach-number shocks form from the vertical and tidal compression of the black hole, the resulting thermal pressure is $\sim 4-6$ orders of magnitude smaller than the ram pressure within the plane, i.e., it is dynamically unimportant and the viscous-less evolution is much closer to reality than when the artificial viscosity is included (as demonstrated directly in \citealt{Kubli:2025}). One could therefore conceive of a spatially dependent viscosity parameterization, such that the $\alpha^{\rm AV}$ and $\beta^{\rm AV}$ terms are ``turned on'' only when the material is near the self-intersection radius, and the inviscid flow is effectively used to establish realistic initial conditions to study the self-intersection process. We defer further investigation of this possibility to future work. 

A second option is to increase the particle number globally (i.e., throughout the duration of the simulation) to the point that the particle density remains high enough near pericenter that spurious viscous forces are absent. The recent work by \citet{Kubli:2025} demonstrates that, upon increasing from $10^6$ to $10^{10}$ particles, the dissipation near pericenter continuously declines to values that are consistent with the expectations from Section \ref{sec:analytic} (see their Figure 3), but a converged amount of dissipation was not reached. Additionally, while such high-particle-number simulations may offer the most accurate and direct means of understanding the disc formation problem in TDEs, it is -- even with high-performance computing and efficient GPU-enabled algorithms -- expensive, and performing, e.g., parameter studies is infeasible. 

A third option, as originally investigated by \citet{Ayal:2000} (see also \citealt{Hu:2026}), is to use a particle-splitting technique, wherein individual SPH particles are divided into lower-mass ``daughter particles'' in under-resolved regions of the flow. Given that the returning flow is hypersonic, if one adopts this methodology and effectively up-samples particle distribution by placing enough fluid elements on neighboring free-particle trajectories, which is obviously accurate in hypersonic and diverging regions of the flow where pressure and viscosity are ignorable by definition, then the velocity field approaches a sufficient level of continuity that its local divergence remains small. 

The issue with this approach is that particle splitting imparts noise to the density distribution that is compounded with each additional level of refinement, as described in e.g.~\citet{Feldman:2007}. If the ratio of the sound crossing time over the stream width to the return time of the debris were small, these splitting-induced perturbations could be smoothed out prior to reaching pericenter. However, the right panel of Figure \ref{fig:tsc} illustrates that this ratio grows and greatly exceeds unity as the fluid elements pass their apocenters, implying that splitting particles upon their ingress will lead to persistent and unphysical noise that negates the possibility of recovering physically robust and numerically consistent levels of dissipation. This noise is apparent from Figure 5 of \citet{Ayal:2000}, who note in the caption of this figure that the clumping is a direct consequence of the splitting method. While it is not as immediately evident because it is a rendering of the density with a restricted range, the bottom-right panel of Figure 2 in \citet{Hu:2026} exhibits similar and noticeable splitting-induced artifacts, and in Appendix \ref{sec:appA} we use the data directly from their simulations (obtained from the public repository \citealt{Hu-Zenodo:2025}) to highlight the presence of these artifacts.  

\begin{figure*}
    \centering
    \includegraphics[width=\textwidth]{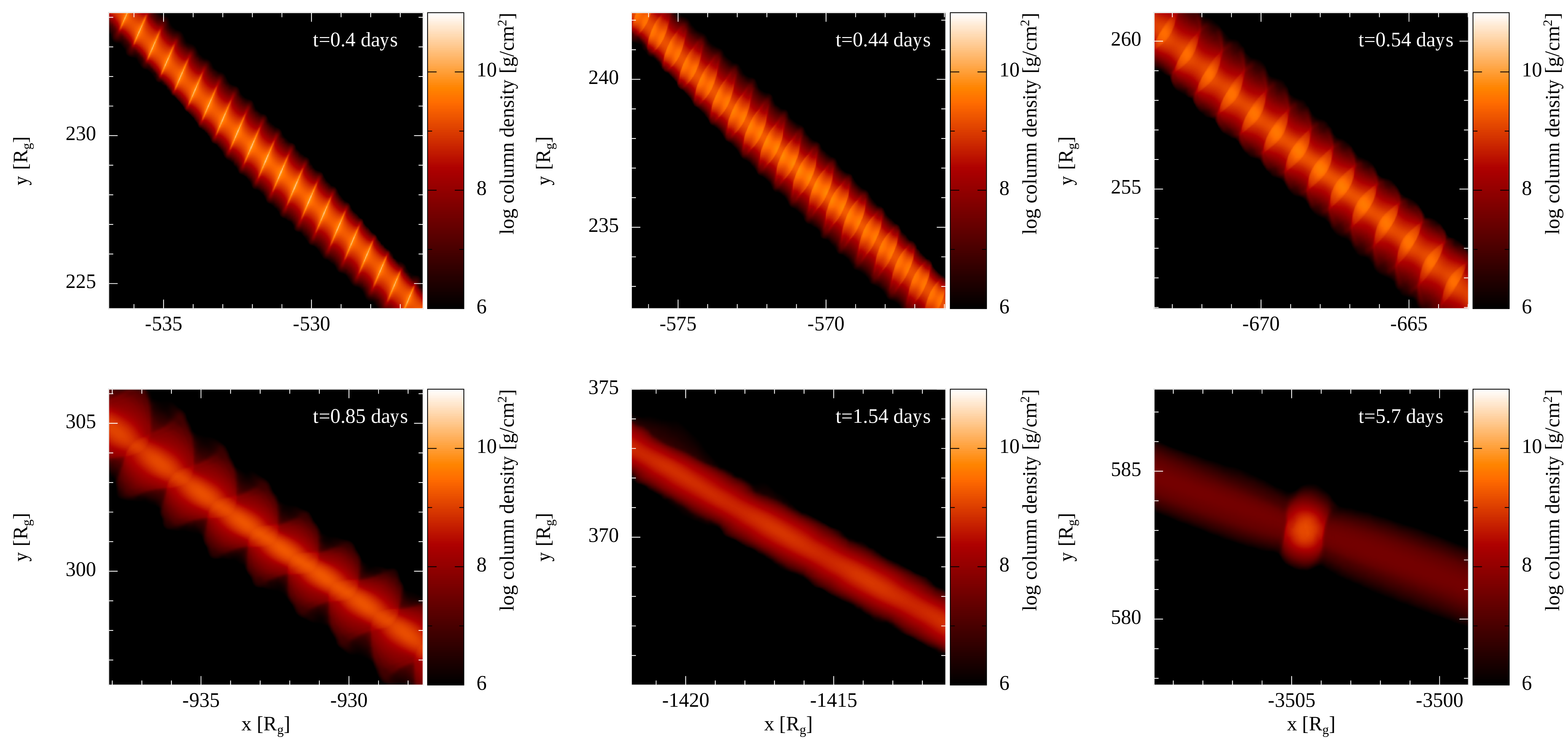}
    \caption{Stream structure at various times following the introduction of a perturbation to the density along the stream before the debris reaches apocentre. Each panel is centred on the same particle in the simulation, and thus the boxes move with the stream. The size of each box in the $x$ and $y$ directions is kept constant, meaning that over time the debris shears out and while many perturbations are present in the first panel, there are less and less as time goes by; with time increasing from left to right and then top to bottom through the six panels. The time of each snapshot is printed on the figure. In this figure we see the induced perturbations propagating, smearing out and leaving a relatively smooth stream. The final panel shows that the perturbations are sufficient to destabilise the stream to the gravitational instability and the stream contains many roughly equally spaced fragments of which one is depicted.} 
    \label{fig:stream1}
\end{figure*}
\begin{figure*}
    \centering
    \includegraphics[width=\textwidth]{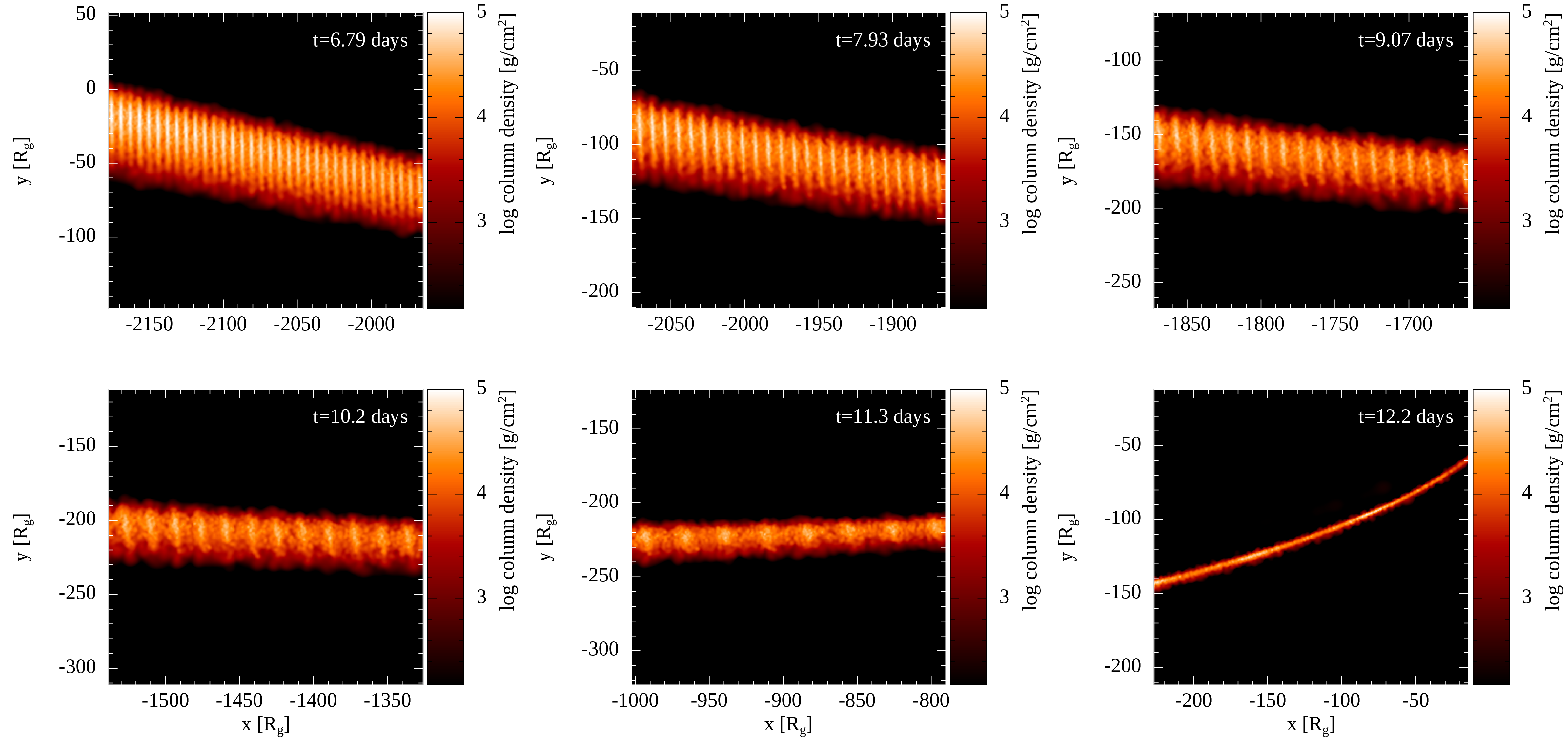}
    \caption{Same as Fig.~\ref{fig:stream1}, except here the perturbation to the density is applied after the debris reaches apocentre. This figures shows that the induced perturbations are not causally connected to each other, despite being spaced much closer than the thickness of the stream, and are sheared out along the stream. The density perturbations are still present as the stream returns to pericentre.}
    \label{fig:stream2}
\end{figure*}

In further support of this conclusion (i.e., in addition to the analytical arguments in Section \ref{sec:analytic} and the artifacts apparent in Figure 5 of \citealt{Ayal:2000} and Figures \ref{fig:panels} \& \ref{fig:dens} below), one can impose noise on the TDE debris stream at various times and simply investigate its evolution. Using the same numerical approach as described in Section \ref{sec:simulations}, but in this case with an $n = 3$ polytrope and a pericenter distance of $r_{\rm t}/2$ to ensure its complete disruption, to this end we performed two tests: one where we imposed a perturbation to the stream at approximately 9.5 hrs (here and below, all times quoted are post-initial-pericentre-passage of the centre of mass of the star), being well before any fluid element has receded to its apocenter, and a second where the perturbation was imposed at a time of 6.8 days, which is well after apocenter for a substantial fraction of the bound material. The perturbation is added by first finding the particles with the largest and smallest radii from the black hole with a density above $0.02$\,g/cm$^3$ ($6\times10^{-6}$\,g/cm$^3$) for the first (second) case, and then using the positions of these particles to define an approximate direction vector, $\boldsymbol{x}_{\rm s}$, along the stream. Each particle above the density cut is then shifted by an amount $\boldsymbol{\Delta x} = A\cos(r_{i}/H)\boldsymbol{x}_{\rm s}$, which effectively introduces a sinusoidal density modulation along the stream, while its velocity is left unchanged. For the first (second) case we take $A = 0.1(1)$, $H = 0.1(1)$. 

The perturbed streams and their evolution are shown in Figures~\ref{fig:stream1} and \ref{fig:stream2}, with each panel of these figures centred on the same particle with the viewing window moving to stay centred on the same patch of fluid. In the first case shown in Fig.~\ref{fig:stream1}, where the perturbation is included before the debris reaches apocentre, the perturbation clearly evolves with time: each density peak spreads, overlaps with its neighbour, and appears to smooth out after a factor of $\sim 5$ in time (bottom middle panel). In the second case shown in Figure~\ref{fig:stream2}, we see that -- even over substantially longer timescales than those depicted in Figure~\ref{fig:stream1} -- the perturbations are not smoothed out. Instead, they are frozen into the flow and stretched and squeezed by the near-ballistic motion of the stream (note that in both figures the size of each box is kept the same, and it is the shear along the stream that reduces the number of peaks and troughs present at later times). We also see that the ratio of the extents of the peaks and troughs along the stream is essentially constant, indicating that there is effectively no communication between them, and by the time the debris reaches pericentre (just after the bottom right panel) the stream still bears the clear imprint of the perturbation induced thousands of gravitational radii away from the black hole.

Figure \ref{fig:stream1} demonstrates a second unintended consequence of seeding the stream with noise: even though the perturbations \emph{appear} to have been smoothed out by the bottom-middle panel, at a later point in time the stream has fragmented into a number of distinct knots, one of which is shown in the bottom-right panel of this figure. This outcome is absent from the perturbation-less stream at the same time, but does occur at a much later time, and is generally the result of its weakly gravitationally unstable nature alongside numerically induced (at the particle level) noise. Specifically, the initial perturbation had a wavelength sufficiently short that it was stable, but as the stream sheared apart the lengthscale of the perturbation increased until it was within the range of unstable wavenumbers for a $5/3$-polytropic cylinder \citep[see][for additional discussion of the gravitational instability of TDE debris streams]{Coughlin:2015, Coughlin:2020, Coughlin:2023}. 

Hydrogen (helium, etc.)~recombination may prevent the stream from fragmenting or at least mitigate the destabilizing influence of self--gravity \citep{Coughlin:2023, Andalman:2025}, but the fact remains that the stream is in a gravitationally precarious state until the debris begins its return to pericenter, and hence particle splitting -- either before or after apocenter is reached -- should not be used as an alternative to achieving higher resolution.

A fourth (and final, at least as concerns this paper) possibility is to use an analytical methodology, such as the one developed in Section \ref{sec:analytic}, \citet{Kochanek:1994}, \citet{Andalman:2025}, etc., to inject SPH particles with accurately known hydrodynamic (and thermodynamic) properties, variants of which have been employed in, e.g., \citet{Jiang:2016, Curd:2021, Bonnerot:2021, Huang:2024}. In this case, the injection site (which could be time-dependent) must be larger than the largest self-intersection radius of the simulated debris. 

The form of Equation \eqref{cd} provides a high value of $\alpha^{\rm AV}$ in regions of the flow that are compression-, rather than shear-, dominated (through the \cite{Balsara:1995} switch, given in Equation \ref{balsara}) and only applied to regions of the flow in which the rate of convergence is increasing (the last term), which indicates a non-linear increase in the density and is typically thought of a precursor to a shock \citep{Cullen:2010}. However, the velocity structure of the debris (Equation \ref{vyeq}) through pericentre passage has a large (positive) value of the time derivative of $-\nabla\cdot\mathbf{v}$ without any corresponding shocks \citep[cf.~Equation \ref{deldotv2} and][]{Kubli:2025}, and the artificial viscosity is therefore being switched on in regions of the flow that do not need it.\footnote{We conjecture that this, i.e.\ the unnecessary application of large artificial viscosity due to the ``detection'' of shocks that are not actually present, is responsible for maintaining the large and anomalous energy spread observed in high-$\beta$ (i.e.\ deep) TDEs even at high resolution \citep{Norman:2021}.} There may therefore be some mileage in employing a shock indicator such as the one proposed by \cite{Rosswog:2020}, which does not depend (at least directly) on the velocity structure of the flow, or one that is built around the existence of shockwaves as a byproduct of the Riemann problem between particles, i.e., Godunov SPH \citep{Inutsuka:1994}.

\begin{figure*}
    \centering
    \includegraphics[width=\textwidth]{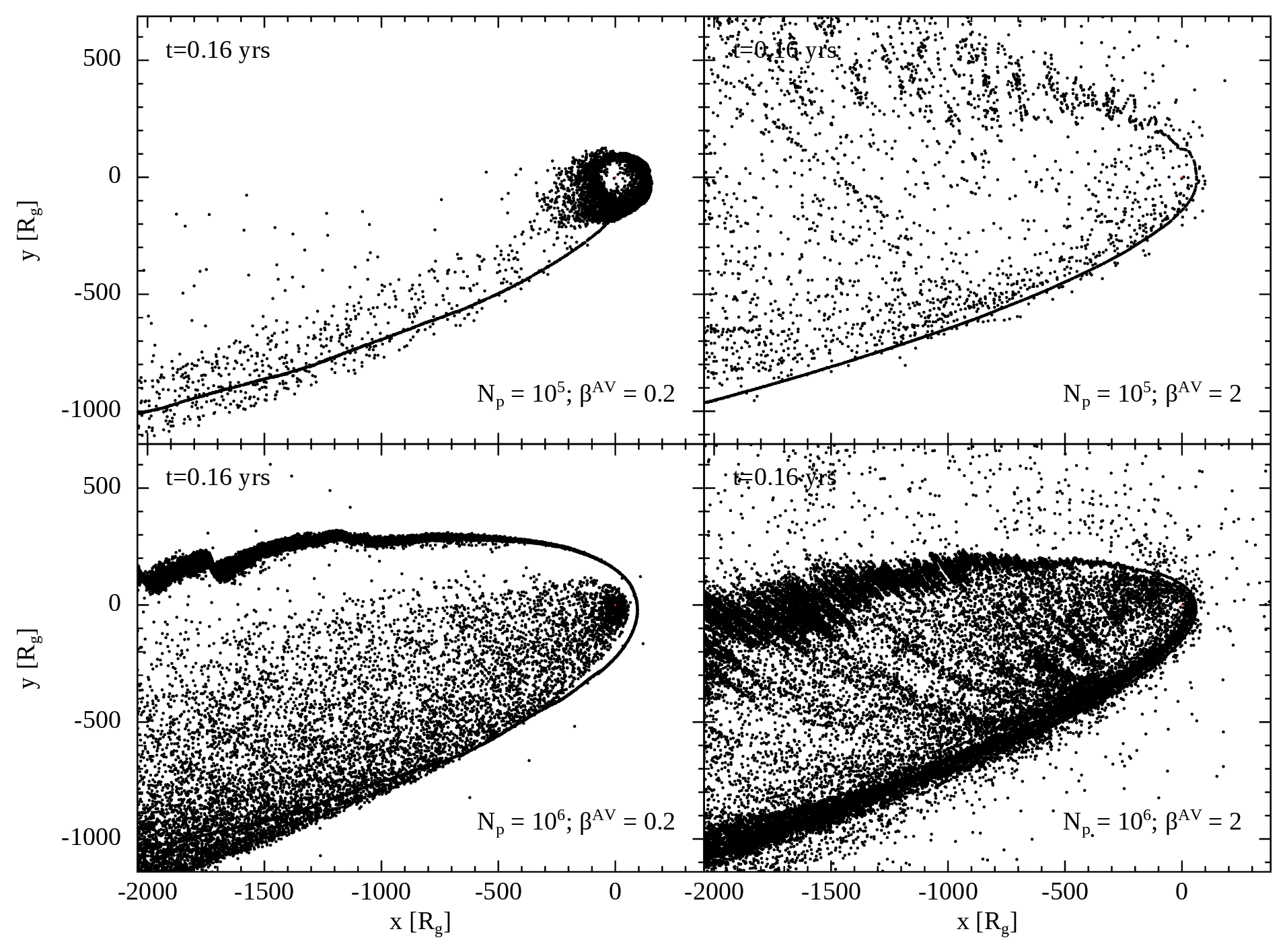}
    \caption{Example debris morphologies showing the range of possible outcomes by varying numerical parameters; top panels show simulations with $10^5$ particles and bottom panels show corresponding simulations with $10^6$ particles, while left panels show simulations with $\beta^{\rm AV} = 0.2$ and right panels show simulations with $\beta^{\rm AV}=2$. In some cases (e.g.~top left) we see rapid and efficient circularisation of the debris into a near-circular disc. In some cases (e.g.~bottom left) there is substantial amounts of direct accretion of high-eccentricity orbits. However, the flow morphology shows substantial dependence on both the artificial viscosity parameter and the resolution employed.}
    \label{fig:discs}
\end{figure*}

Finally, in Figure \ref{fig:discs} we show that some of the simulations (without heating from artificial viscosity) lead to efficient disc formation. The number of particles used and the $\beta^{\rm AV}$ viscosity coefficient are given in the panel legend. The top panels show the $\beta^{\rm AV} = 0.2$ and $\beta^{\rm AV} = 2$ cases with $10^5$ particles. For $\beta^{\rm AV} = 2$, and also $\beta^{\rm AV}=0.02$ (not shown), the stream returning to pericentre does not maintain a coherent structure and emerges in a wide-angle fan. However, $\beta^{\rm AV}=0.2$ appears to represent a ``sweet spot'' where the breadth of the fan is sufficiently small, but the interaction with the incoming stream sufficiently strong, that the debris rapidly forms a small-scale disc at roughly the circularisation radius of the debris. Conversely in the lower panels -- which are identical to the top panels except in particle number -- the material in the $\beta^{\rm AV} = 0.2$ simulation is more directly aimed at the black hole, with a significant amount of material directly swallowed on high-eccentricity orbits and a fraction of it forming a smaller-scale (i.e., smaller than the circularisation radius of the initial stream) disc. Thus, with suitable choice of the artificial viscosity parameters and particle number (resolution), we can obtain (1) rapid circularisation into a disc, (2) direct accretion of high-energy orbits with little energy released, or (3) delayed accretion with material expelled in a low-density, hot fan/bubble. Unfortunately, not one of these exciting results 
stands up to the scrutiny of convergence testing. We are therefore seemingly no closer to determining the amounts of stellar debris that accrete through a disc, accrete from high eccentricity orbits, or escape in a wind, as these fractions depend on numerical parameters; the various possibilities spelled out in \citet{Rees:1988} remain.

\section{Summary}
\label{sec:summary}
With the overarching theme of establishing a frank and open narrative of some of the issues in modelling TDE debris flows with 3D numerical hydrodynamical algorithms, in this paper we have (1) provided a relatively simple analytical model that enables an understanding of the velocity and thermodynamic structure of the TDE debris stream as it returns to pericenter (Section \ref{sec:analytic}); (2) highlighted the inherent difficulties presented by this structure in terms of often-adopted numerical hydrodynamical algorithms, which leads to the numerical nature of some of the properties of TDE flows present in the literature (Section \ref{sec:Origin}), and (3) elucidated the ways in which the numerical methodology can be improved to enable low-cost, high-accuracy simulations (Section \ref{sec:discussion}). While we have focussed on the SPH method, we have tried to make it clear throughout that the discussion also applies to grid-based codes: while modern finite-volume codes do not typically employ explicit viscous dissipation, the shocks that form between underresolved grid cells as a byproduct of the Riemann problem effect the same outcome. We emphasize that our motivation and aim for this discussion is not to denigrate the methods themselves, or to imply that they cannot be used to study these phenomena (indeed, \citealt{Kubli:2025} have shown that with sufficient resolution the flow structure can be accurately modelled with 3D SPH), but to explore where and how the methods can be improved. We hope we have achieved these aims.

\section*{Acknowledgments}
CJN acknowledges support from the Leverhulme Trust (grant No. RPG-2021-380) and from the Science and Technology Facilities Council (grant No. ST/Y000544/1). ERC acknowledges support from the National Aeronautics and Space Administration through the Astrophysics Theory Program, grant 80NSSC24K0897. ZLA was supported by the U.S. Department of Energy, Office of Science, Office of Advanced Scientific Computing Research, Department of Energy Computational Science Graduate Fellowship under Award Number DE-SC0024386. This report was prepared as an account of work sponsored by an agency of the United States Government. Neither the United States Government nor any agency thereof, nor any of their employees, makes any warranty, express or implied, or assumes any legal liability or responsibility for the accuracy, completeness, or usefulness of any information, apparatus, product, or process disclosed, or represents that its use would not infringe privately owned rights. Reference herein to any specific commercial product, process, or service by trade name, trademark, manufacturer, or otherwise does not necessarily constitute or imply its endorsement, recommendation, or favoring by the United States Government or any agency thereof. The views and opinions of authors expressed herein do not necessarily state or reflect those of the United States Government or any agency thereof. We used {\sc splash} \citep{Price:2007} for some of the figures. Some of this work was undertaken on the Aire HPC system at the University of Leeds, UK. Some of this work used the DiRAC Data Intensive service DIaL3 at the University of Leicester, managed by the University of Leicester Research Computing Service on behalf of the STFC DiRAC HPC Facility (www.dirac.ac.uk). The DiRAC service at Leicester was funded by BEIS, UKRI and STFC capital funding and STFC operations grants. DiRAC is part of the UKRI Digital Research Infrastructure.




\bibliographystyle{aasjournal}
\bibliography{nixon} 

\appendix

\label{sec:appA}
First used by \citet*{Ayal:2000} in the context of TDEs, \citet*{Hu:2026} re-used the technique of particle splitting\footnote{Specifically, they used the ``Adaptive Particle Refinement'' algorithm (previously known as particle splitting) reintroduced by \cite{Nealon:2025}.} to understand the dissipation associated with the debris as it returns to pericenter, arguing that by ``shuffling'' the particles they could reduce the noise it otherwise introduces. They also stated that the distance between each spatial region in which the stream was split (see their Figure~1) enabled ``$\geq 100$ sound crossings of the gas, allowing particles to fully relax between splitting and merging events and reducing the total noise.'' This contradicts the estimates in Section \ref{sec:analytic} shown in Figure \ref{fig:tsc}, i.e., according to our analytical model it should be out of causal contact and any perturbations to the stream should remain. The predictions of our analytical model were upheld by simulations in which we included a perturbation to the stream (Figures~\ref{fig:stream1} \& \ref{fig:stream2}). 

\begin{figure*}
    \includegraphics[width=0.995\textwidth]{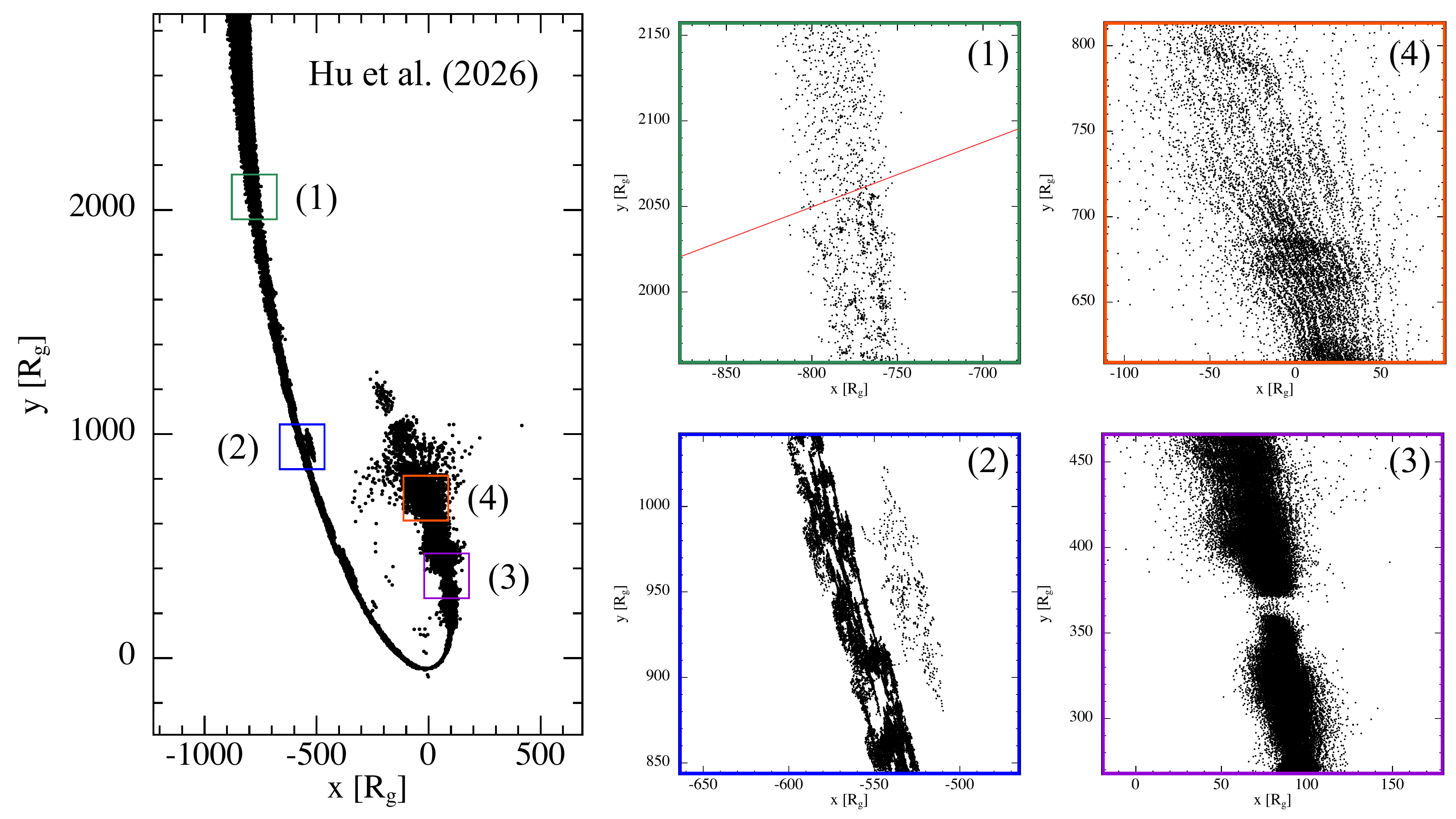}
    \caption{Particle distribution plots of the debris stream from the simulation presented in \protect\cite*{Hu:2026}. The left panel shows the main structure of the stream; there is a noticeable fanning of the stream exiting pericentre. There are, however, additional anomalous features present that are highlighted by the four coloured boxes, which correspond to the zoom-in panels on the right. The zoom-ins have coloured borders that match the coloured squares in the left hand panel (also marked by the numbers). Panel (1) contains a red line that marks the location of the first splitting point -- particles at radii larger than this line do not experience splitting -- in the simulation presented by \protect\cite*{Hu:2026}, and a transition from approximately smooth flow above to clumpy flow below this line can easily be seen. Panels (2)-(4) contains additional features that have no physical origin.}
    \label{fig:panels}
\end{figure*}

To investigate whether the artifacts from particle splitting are indeed preserved, or whether they are somehow sufficiently smoothed out by the shuffling procedure employed by \citet*{Hu:2026}, we have explored the particle distribution in the data taken directly from the simulation presented in \citet*{Hu:2026} provided by \cite{Hu-Zenodo:2025}. Figure \ref{fig:panels} shows the raw particle distribution corresponding to the lower right panel (16 levels of refinement) in Figure 2 of \cite*{Hu:2026}, accompanied by four separate zoom-ins on the distribution at various locations in the stream. The first zoom-in (top-left, bounded by green) encompasses the first splitting region (demarcated by the red-line, i.e., particles at radii larger than the red line have not been split, while those interior to it have), the second (bottom-left, bounded by blue) is midway towards pericentre, the third (bottom-right, bounded by purple) is just after pericentre and the fourth (top-right, bounded by orange) encloses material that has travelled further away from pericentre. The stream, from the first splitting of particles, takes on an incorrigible state\footnote{i.e.~beyond correction or reform.} and each successive splitting compounds the issue. This plot also makes it clear that---while not visible in the corresponding density rendering provided in Figure 2 of \citet*{Hu:2026}---the outgoing stream is considerably wider after passing through pericenter.

\begin{figure}
    \includegraphics[width=0.495\textwidth]{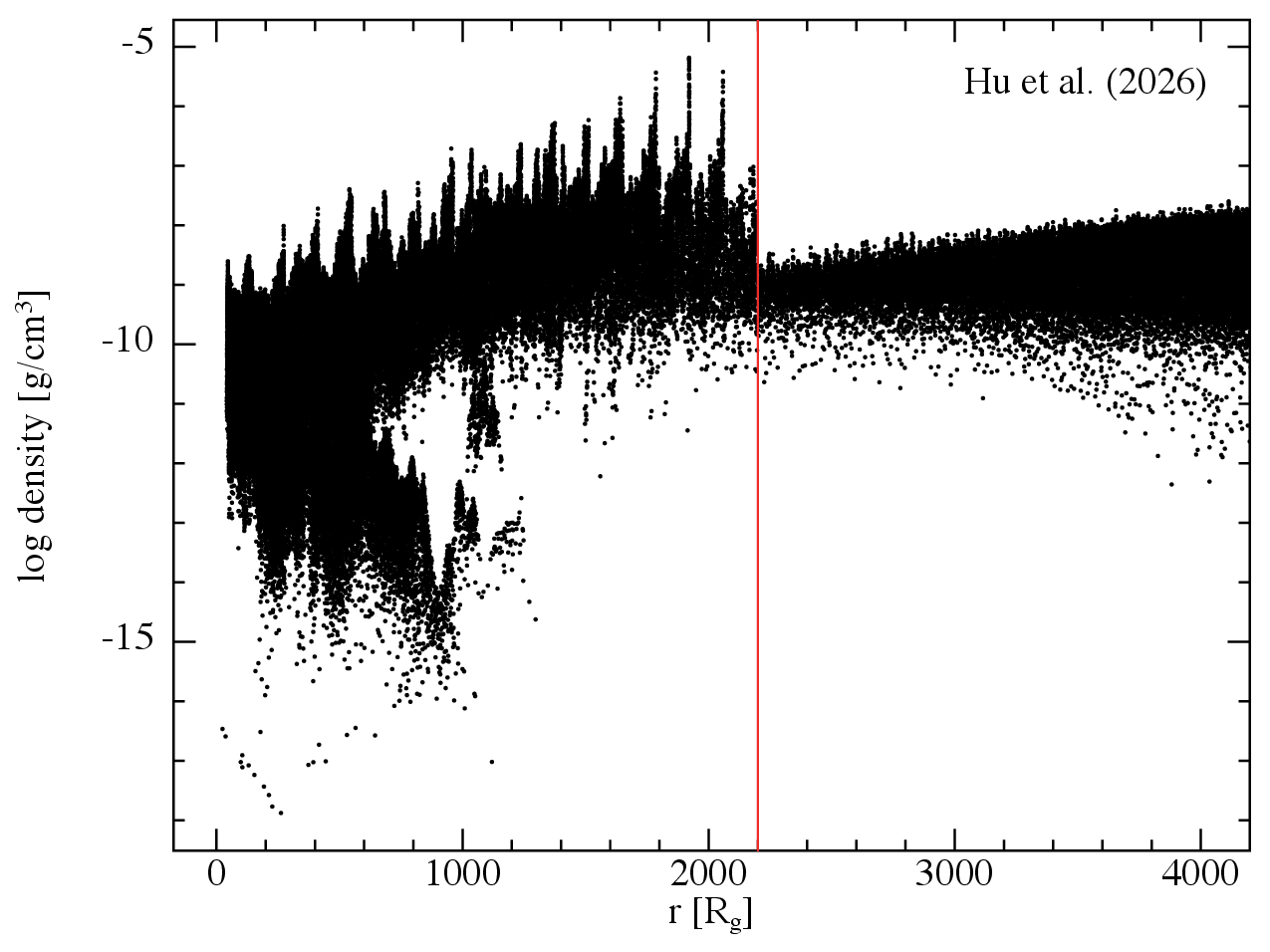}
    \includegraphics[width=0.2475\textwidth]{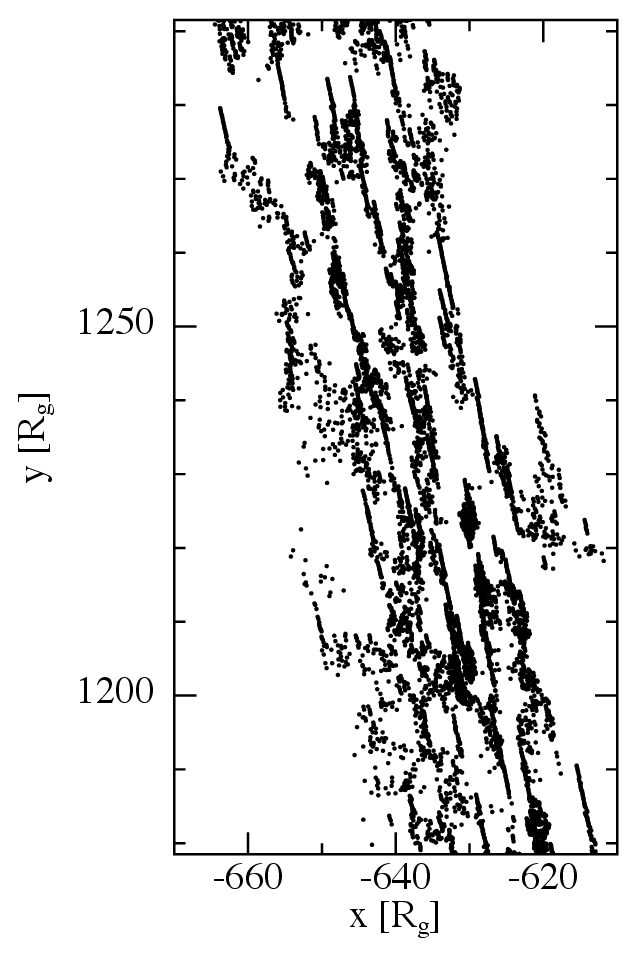}
    \includegraphics[width=0.2475\textwidth]{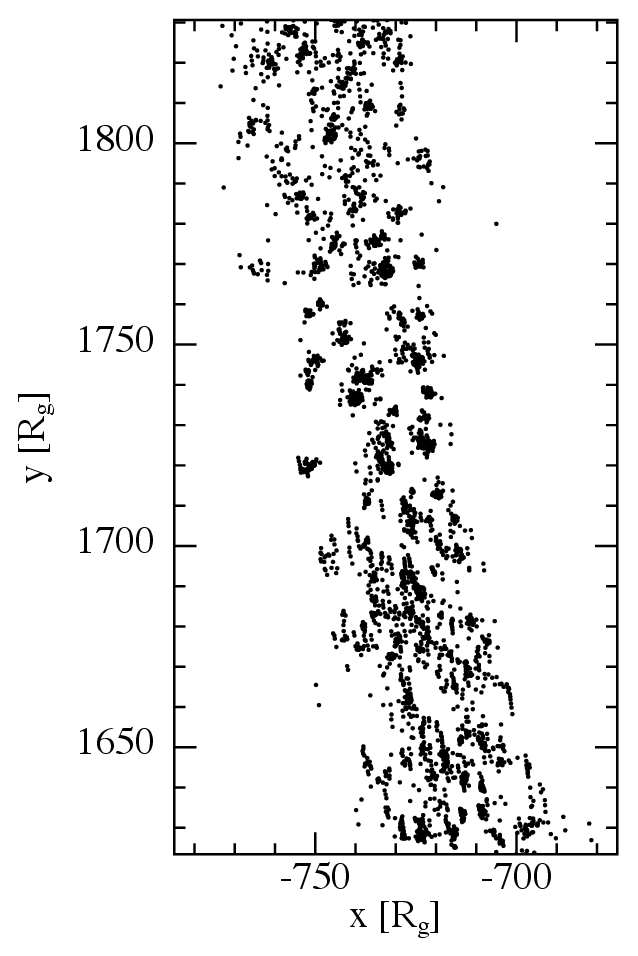}
    \caption{Left panel: The density of particles plotted against radius in the simulation presented in \protect\cite{Hu:2026}. The red line marks the location of the first particle-splitting region (with the stream moving from right to left on the plot) and coincides with an instantaneous and large jump in the densities of the particles that is typical of a noisy, artificially-fragmented stream. Middle and right panels: Zoom-ins on the particle structure in the stream. The lines (middle panel) and knots (right panel) of particles are the patently unphysical features responsible for the substantial density increases seen in the left panel.}
    \label{fig:dens}
\end{figure}

Additionally, we show in Figure~\ref{fig:dens} (left panel) the density of the particles in the simulation presented by \cite*{Hu:2026} as a function of their distance from the black hole. The red line marks the location of the outermost splitting zone, which is coincident with the red line in the first zoom-in in Figure \ref{fig:panels}, with 15 more zones inside this. The noise in the particle distribution, introduced by the splitting procedure, results in density variations that are up to 4 orders of magnitude above that of the unsplit stream (see the particle structures shown in the middle and right panels of Figure~\ref{fig:dens}). Finally, it is worth noting that the particle structures developed in the simulations presented in \cite*{Hu:2026}, and shown here in detail (Figures~\ref{fig:panels} \& \ref{fig:dens}), do not occur in simulations performed with codes that employ a robust SPH algorithm such as the one implemented in {\sc phantom} without particle splitting \citep[cf.~Section 2.4.1 of][]{Price:2012b}.

\end{document}